\documentclass[twocolumn,dvipsnames]{aastex63}
\RequirePackage{lineno}
\usepackage{graphicx}	
\usepackage{multirow,makecell}

\RequirePackage{bm,amsfonts,hhline,amsmath,amssymb,microtype,float,eurosym,
latexsym,epsf,mathtools,cuted,times,makecell,array,natbib,cancel}
\usepackage{tabularx}
\usepackage{epstopdf}
\usepackage[normalem]{ulem}
\usepackage{gensymb}
\usepackage{xcolor}

\usepackage{pifont}

\usepackage{hyperref}

\hypersetup{colorlinks=true, linkcolor=red, filecolor=magenta, urlcolor=Sepia, citecolor=blue}

\definecolor{ForestGreen}{HTML}{228B22}

\newcommand{\mpla}{MPLA}
\newcommand{\ijmpd}{IJMPD}
\newcommand{\aop}{AnPhy}
\newcommand{\pr}{PhRv}
\newcommand{\plb}{PhLB}

\received{2021 October 24}
\revised{2021 December 2}
\accepted{2021 December 6}

\submitjournal{ApJ}

\shortauthors{Deb, Mukhopadhyay \& Weber}
\shorttitle{Anisotropic magnetized white dwarfs}

\begin{document}

\title[Anisotropic magnetized white dwarfs]{Anisotropic magnetized
  white dwarfs: Unifying under- and over-luminous peculiar and
  standard type Ia supernovae}

\author[0000-0003-4067-5283]{Debabrata Deb$^{\dagger}$}
\email{debabratadeb@iisc.ac.in$^{\dagger}$} \affiliation{Department of
  Physics, Indian Institute of Science, Bangalore 560012, India}

\author[0000-0002-3020-9513]{Banibrata Mukhopadhyay$^{\ddagger}$}
\email{bm@iisc.ac.in$^{\ddagger}$} \affiliation{Department of Physics,
  Indian Institute of Science, Bangalore 560012, India}

\author[0000-0002-5020-1906]{Fridolin Weber$^{\star}$}
\email{fweber@sdsu.edu$^{\star}$} \affiliation{Department of Physics,
  San Diego State University, San Diego, CA 92182, USA}
\affiliation{Center for Astrophysics and Space Sciences, University of
  California at San Diego, La Jolla, CA 92093, USA}



\begin{abstract}
 Ever since the observation of peculiar over-luminous type Ia supernovae (SNeIa), exploring possible violations of the  canonical Chandrasekhar mass-limit (CML) has become a pressing research  area of modern astrophysics. Since its first detection in 2003, more than a dozen of peculiar over-luminous SNeIa has been detected, but the true nature of the underlying progenitors is still under dispute.  Furthermore there are also under-luminous SNeIa whose progenitor masses appear to be well below the CML (sub-Chandrasekhar progenitors). These observations call into question how sacrosanct the CML is.  We have shown recently in Paper I that the presence of a strong magnetic field, the anisotropy of dense matter, as well as the orientation of the magnetic field itself significantly influence the properties of neutron and quark stars. Here,  we study these effects for white dwarfs, which shows  that their properties are also severely impacted.  Most importantly, we arrive at a variety of mass--radius relations of white dwarfs which 
accommodate sub- to super-Chandrasekhar mass limits.  This urges caution when using white dwarfs associated with SNeIa as standard candles. \\ 

{\it Unified Astronomy Thesaurus concepts:} White dwarf stars (1799); Chandrasekhar limit (221); 
Type Ia supernovae (1728); Degenerate matter (367); Magnetic fields (994); Massive stars (732); Magnetic stars (995); 
Gravitation (661)\\
\end{abstract}




\section{Introduction}\label{sec:intro} 
Recently,~\cite{DMW} initiated the exploration of effects of anisotropy on strongly magnetized compact stars in general relativity. They showed in their work (hereinafter referred to as Paper I) that not
only the presence of a strong magnetic field and anisotropy, but also the orientation of the magnetic field itself, significantly influence the physical properties of stars.  They further showed that static equilibrium is not achieved, unless both the local matter anisotropy effects and the anisotropy effects caused by a strong magnetic field are considered. Paper I showed that for a transversely oriented magnetic field, the stellar mass increases with increasing field strength, which is however opposite to a radially oriented field. Based on a detailed exploration, it was argued that massive neutron and strange quark stars of mass more than $2.5M_\odot$ are possible. The authors also briefly touched upon the possible violation of the Chandrasekhar mass-limit (CML), based on this idea. 
\cite{Chandrasekhar1931,Chandrasekhar1935} first introduced the idea of the possible limiting mass for WDs in his celebrated works, where he predicted that beyond $\sim 1.4\, M_\odot$ no non-rotating and non-magnetized carbon-oxygen (C-O) WD exists. This critical mass limit is famously known as the CML for WDs.

In this article, we explore in detail the effects of the above
mentioned anisotropy due to high magnetic field strengths as well as  matter properties in white dwarfs (WDs).
Stable equilibrium configurations of WDs are achieved due to the repulsive electron degeneracy pressure which counterbalances the inward gravitational pull. When WDs gain mass (say, via mass accretion from companion stars) and exceed a certain critical mass limit, the inward gravitational force overpowers the repulsive force of the degenerate electrons, which leads to a situation where fusion reactions are initiated. Within a few seconds,  runaway reactions set in to unbind a considerable amount of WD matter through an explosion which releases an enormous amount of energy $\sim{10}^{51}$ erg, known as a type Ia supernova (SNIa). This violent cosmic event ensures 
generally complete disruption of WDs without leaving behind any stellar remnant.
However, recently another class of supernovae related to WDs, named
type Iax supernovae, has been observed which look similar to SNeIa but
are much fainter.  In the case of type Iax supernovae, the progenitors may
partially survive the explosions and move away at high kick
velocities (\citealt{Vennes2017}). 

Type Ia supernovae (SNeIa) manifest a specific set of relations between intrinsic luminosity, colour, and light-curve width~\citep{Phillips1993,Goldhaber2001}. Such standard and stable physical features of SNeIa help astronomers to use them as ``\emph{standard candles}'' to accurately measure the distances to their host galaxies.  The variation of the brightness of SNeIa with the distance is a key technique used to measure cosmological parameters. In fact, the pioneering observation of the accelerated expansion of the Universe, which confirms the probable role of dark energy behind this cosmic phenomenon, was inferred from the consideration of SNIa to be a {\emph{``standard candle''}}~\citep{Riess1998,Perlmutter1999}. However, in recent years, a series of observations of several peculiar over-luminous SNeIa, such as \emph{SN 2003cv}~\citep{Howell2006}, \emph{SN 2007fg}~\citep{Scalzo2010}, \emph{SN 2009if}~\citep{Taubenberger2011} and \emph{SN 2013dc}~\citep{Cao2016},  constitute a serious setback to the generally accepted \emph{``standard candle''} concept,  as they are best explained by massive progenitor WDs having mass beyond the standard CML. As SNeIa are mainly powered by the radioactive decay of $^{56}\textrm{Ni}$, to explain 2.2 times overluminous SN 2003cv,~\cite{Howell2006} predicted that it requires $\sim 1.3\, M_{\odot}$ of $^{56}\textrm{Ni}$ which leads to a possible progenitor WD of mass $\sim 2.1\, M_{\odot}$, popularly known as super-Chandrasekhar progenitor WD (SCPWD). Later,  researchers discovered the overluminous SN 2009if having high mass SCPWD as $\sim2.8\, M_\odot$.

Due to peculiar features of these overluminous SNeIa, they immediately attracted the attention 
of researchers who came up with two possible explanations such as (i) mergers of massive WDs~\citep{Iben1984,Hicken2007} and (ii) explosions of rapidly rotating WDs~\citep{Howell2006,Boshkayev2013,Branch2006}. In a double C-O WD system, one star accretes mass from the companion C-O rich WD. In the accreting WD, the ignition of carbon burning starts due to heating of the outer layer if the accretion rate exceeds the critical limit of $2.7\times {10}^{-6}\, M_\odot \textrm{yr}^{-1}$~\citep{Nomoto1985,Kawai1987}. The final stage of the accreting WD depends on whether carbon ignition is at the center or in the envelope of the star. If carbon ignition starts at the stellar core of a WD, it leads to a SNIa explosion. On the other hand, ignition in the stellar envelope leads to the propagation of a deflagration flame towards the center and the immediate conversion of a C-O WD to an O-Ne-Mg WD~\citep{Saio1985,Saio1998,Timmes1994}, which finally collapses into an ultra-dense neutron star~\citep{Nomoto1984,Nomoto1987,Nomoto1991}.  In both cases, the double degenerate scenario does not lead to a SNIa explosion that could explain $2.8~M_{\odot}$ SCPWD. In fact,  numerical simulations of massive WD mergers indicate the collapse of an accreting WD to a neutron star due to off-center carbon burning~\citep{Saio2004,Martin2006}. On the other hand,~\cite{Chen2009} showed that a differentially rotating accreting WD in close vicinity to its companion star can not exceed a mass of $~1.7\,M_\odot$.  Therefore, it is clear that the above discussed two competing scenarios of SNIa progenitors fail to explain  high-mass SCPWDs.  

Later, Mukhopadhyay and collaborators~\citep{Das2012a,Kundu2012} introduced a magnetic WD model where they suitably explained SCPWDs. This model considers high magnetic fields within the stellar structure having Landau levels in the plane perpendicular to the axis of the magnetic field and is therefore able to explain highly massive SCPWDs with masses of~$\sim 2.6\,M_\odot$~\citep{Das2013a}.~\cite{Das2014a} also proposed that if the strong magnetic field within the magnetized WDs (hereinafter B-WDs) is fluctuating/tangled at a length scale larger than the quantum length scale, the average magnetic field and corresponding magnetic pressure could be considerably smaller than the matter pressure, which is however modified by the Landau quantization. The degree of Landau quantization influences the quasi-equilibrium state of highly magnetized B-WDs, which modifies the matter pressure that counterbalances the inward gravitational pull. Hence, the mass--radius relation of B-WDs deviates more and more from the CML the longer accretion of mass continues~\citep{Das2013a,Das2013b}. The mass loading onto B-WDs leads to an increase in central density which contracts their size~\citep{Cumming2002}. As a consequence, the central magnetic field of B-WDs may increase accordingly due to the magnetic flux freezing theorem.

Interestingly, by applying the varying accretion rate scenario,~\cite{Das2013b} argued that $0.2\, M_\odot$ WD with $10^9$~G surface magnetic field ($B_s$) turns into a super-Chandrasekhar B-WD (SCBWD) within $2\times {10}^{7}$ years. Mukhopadhyay and his group over the years have been attempting to develop a more sophisticated and generalized model that would explain the complex astrophysical situations for SCBWDs realistically by accounting for general relativistic effect,  differential rotation, variations of the magnetic fields and  geometry,  Landau quantization, thermal luminosity, deviations from spherical symmetry, etc.,  see~\citealt{Das2014b,Das2015a,Subramanian2015,Mukhopadhyay2016,Mukhopadhyay2017,Bhattacharya2018,Gupta2020,Mukhopadhyay2021}. They~\citep{Kalita2019,Kalita2020} also showed how it is possible to detect SCPWDs directly via continuous gravitational wave astronomy in the various upcoming space-based detectors, such as DECIGO, LISA and BBO. The idea of SCBWDs proposed by Mukhopadhyay and his group has also been explored and confirmed by different independent research works, such as by~\cite{Federbush2015,Franzon2015,Franzon2017,Moussa2017,Shah2017,Sotani2017,Roy2019}.

As described in Paper I, \cite{Bowers1974} strongly argued against the over-simplistic assumption that compact stars are entirely made up of an isotropic perfect fluid. They generalized the Tolman-Oppenheimer-Volkoff (TOV) equation~\citep{Tolman1939,Oppenheimer1939}, which describes the structure of spherically symmetric stars, to systems where pressure is anisotropic. They also showed that the inclusion of local anisotropy leads to  non-negligible effects on the properties of stars, such as the total mass, radius, density, and surface redshift.  Interestingly, the spontaneous creation of strong magnetic fields within a compact star breaks the spatial rotational symmetry~$\mathcal{O}\left(3\right)$, which leads to pressure anisotropy within the stellar structure~\citep{Ferrer2010,Isayev2011,Isayev2012}. As to the other origin of anisotropy, it is the consequence of cooling process of WDs which takes almost the entire time of its evolutionary stage, lasting for nearly $10$~Gyr. During the cooling at low temperatures, a first-order fluid-solid transition occurs in the WD's dense plasma, as accurately predicted by~\cite{Kirzhnits1960,Abrikosov1960,Salpeter1961} in the early sixties. Later,~\cite{Stevenson1980} showed via his proposed phase diagram that carbon and oxygen are immiscible in  the solid phase, which was confirmed by~\cite{Garcia1988}.~\cite{Nag2000} in their study discussed the cooling of WDs via condensation of a Bose gas leading to a first-order phase transition. They also predicted that the cores of  massive WDs are actually Bose condensed and made of ``\emph{crystalline normal crustal matter}", which is comparable to the ultra-dense neutron star structure, including ``\emph{superfluid core and normal neutron matter crust}". 

While one may find several works treating magnetized nonspherical stars in general relativity with publicly available numerical codes (XNS and Lorene) or the perturbative approach solving the Einstein-Maxwell field equations, with toroidal or poloidal or mixed field geometries, (see,
e.g.,~\citealt{Bocquet1995,Konno1999,Cardall2001,Ioka2004,Oron2002,Kiuchi2008,Kiuchi2009,Yasutake2010,Frieben2012,Yoshida2012,Das2015a,Subramanian2015,Kalita2019,Kalita2020}),
none of them include the effects of anisotropy, making them incomplete. Importantly, in Paper I we showed that it is essential to consider the effective anisotropy,  both due to the fluid and the field,  as it has a notable impact on the structure of magnetized stars. Therefore, in the light of the above discussions, in this work, we include the pressure anisotropy in order to obtain a generalized picture of the structure of WDs.

Recently,~\cite{Chowdhury2019} proposed an anisotropic spherically symmetric model for non-magnetized WDs in the framework of scalar-tensor (ST) theories of gravity. On the other hand,~\cite{Chu2014} showed that the magnetic field strength and its orientation significantly affect the maximum mass of stars. Although they attempted to consider magnetic field orientations, the effects of the magnetic field orientation and of the pressure anisotropy are not explicitly included in the TOV equation used in their model, which was rectified in Paper I. Hence, in the present work, we include both the magnetic field strength and its orientation in the TOV equation to achieve a more complete description of the effects of magnetic fields and anisotropy on the properties of WDs. Like in Paper I, we denote the orientation of the magnetic fields along the radial direction as ``Radial Orientation (RO)'' and their orientation along the orthogonal to the radial direction (say along the $\theta$- or $\phi$-directions) as ``Transverse Orientation (TO)''. Importantly,  we further show that stability of B-WDs is not  achieved unless we consider anisotropy of the system arising from the combined effects of (i) anisotropy due to strong magnetic fields and (ii) anisotropy of the system fluid. 

We organize this Paper as follows. In Section~\ref{sec1} we discuss the basic formalisms for B-WDs and show that it is important to consider anisotropy due to both the fluid and field. In Section~\ref{sec2} we discuss the results and show the effects of anisotropy, magnetic fields and their orientations on the different physical properties of WDs. Important concluding remarks are provided in Section~\ref{sec3}.


\begin{figure*}
\centering
   \label{density_WD1} \includegraphics[width=.33\textwidth]{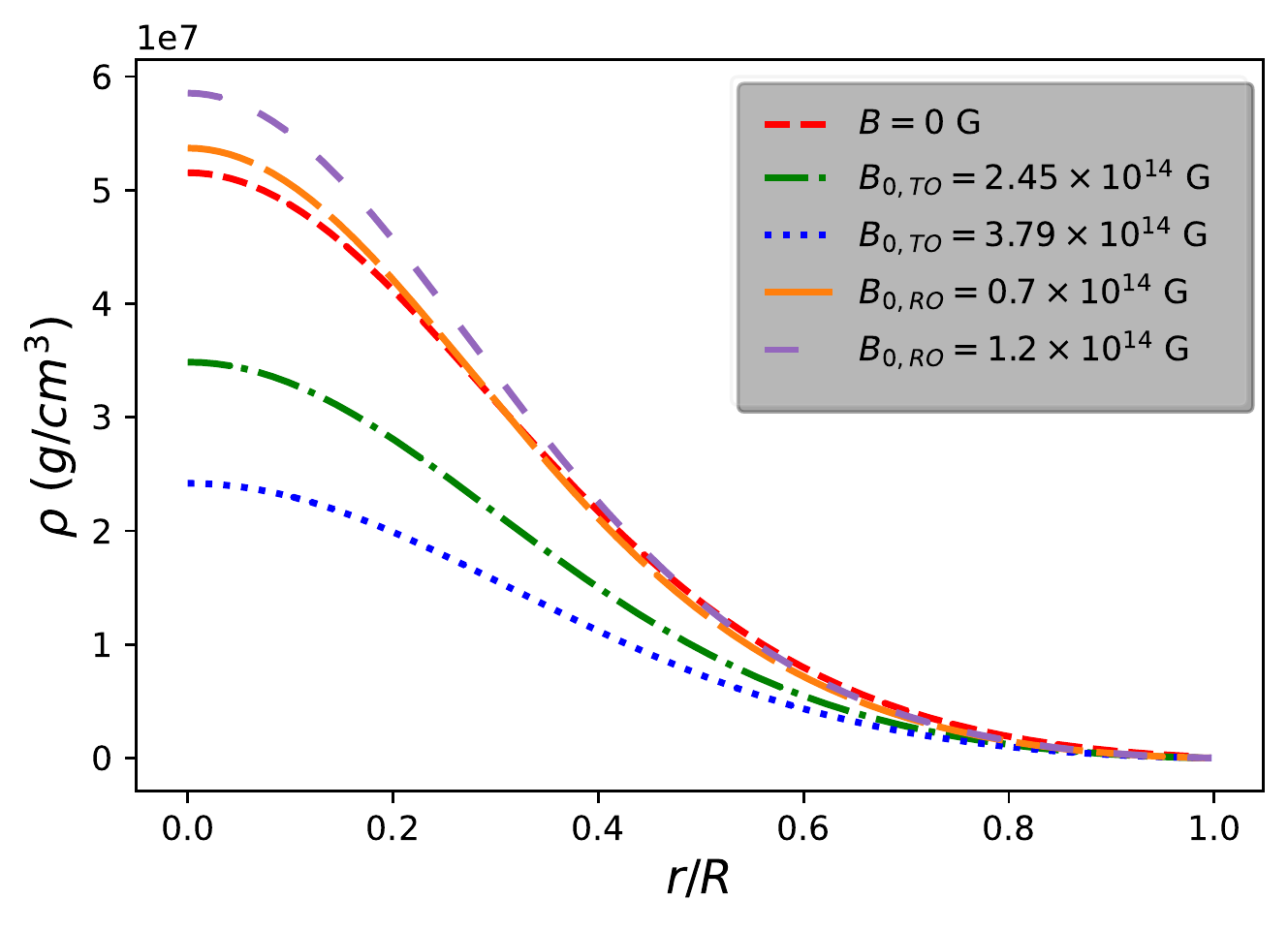}\hfill
   \label{radpress_WD1} \includegraphics[width=.33\textwidth]{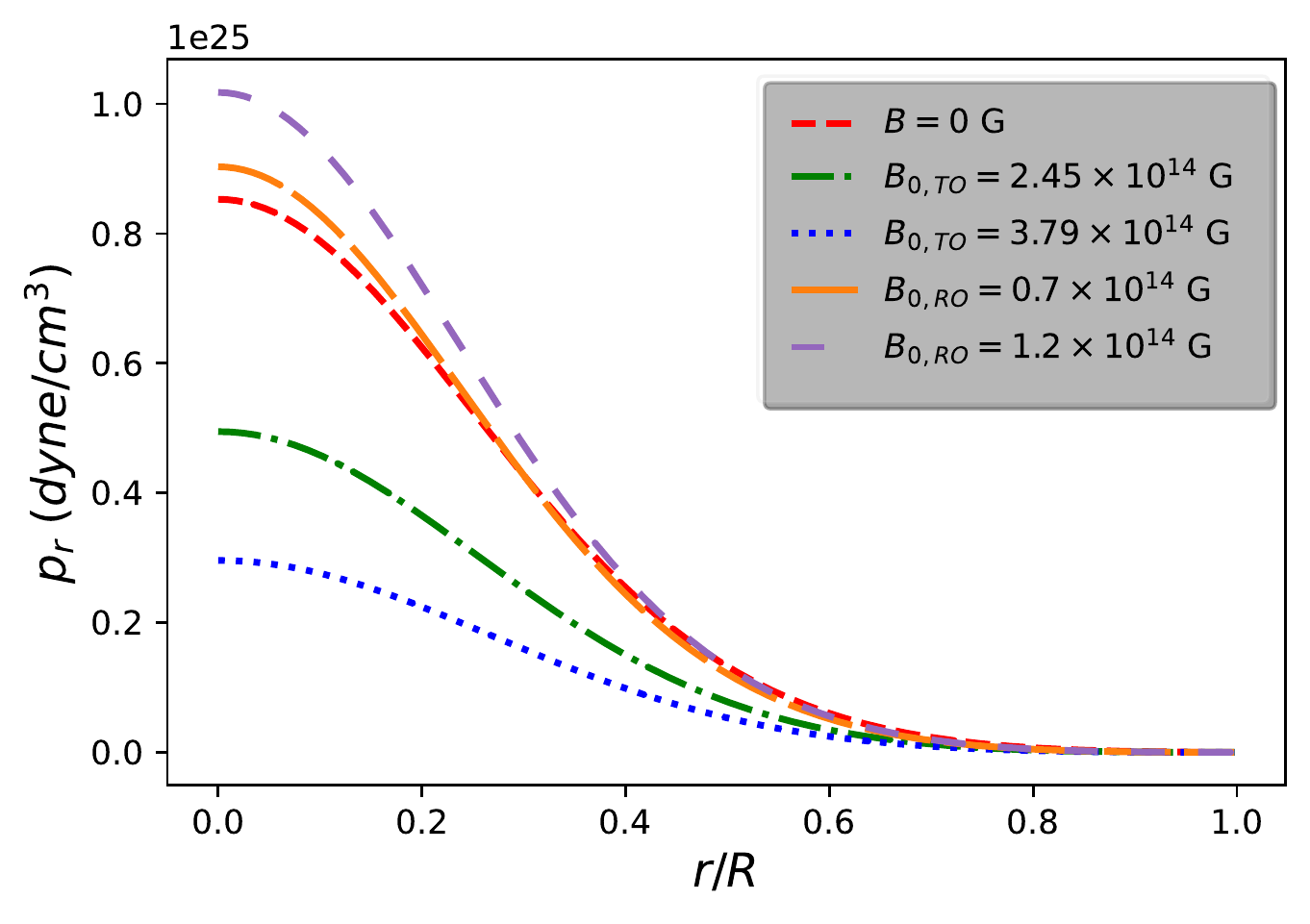}\hfill
   \label{tangpress_WD1}\includegraphics[width=.33\textwidth]{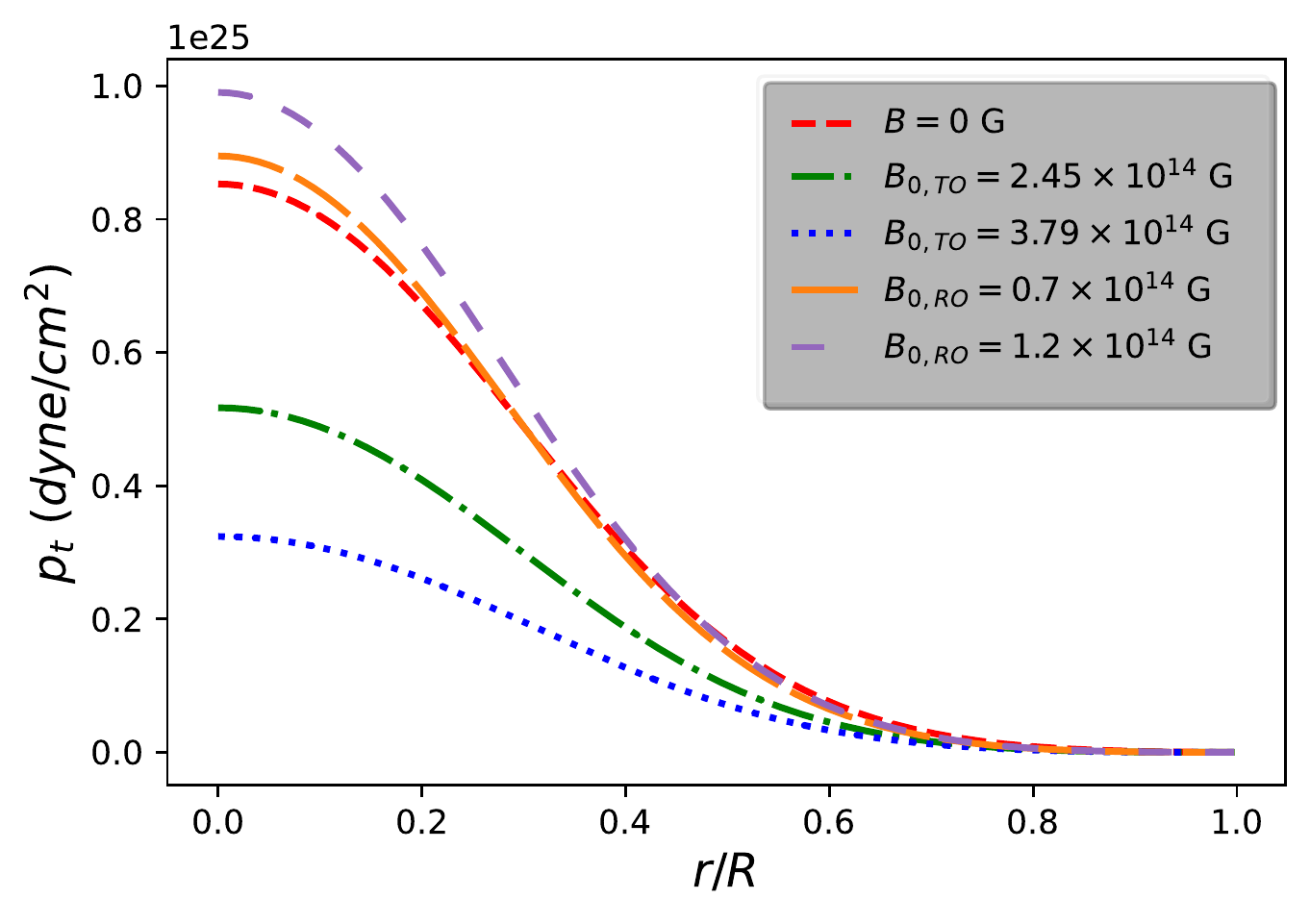}
	\caption{Variation of (a) matter density ($\rho$) (left panel), (b) radial pressure $(p_r)$ (middle panel) and (c) tangential pressure $(p_t)$ (right panel) with radial distance $r/R$ of a magnetized $1.3\, M_\odot$ WD.  Here and in what follows $\kappa=0.45$, $\eta=0.2$, $\gamma=0.9$ and $B_s = {10}^9$~G. The labels `RO' and `TO' are explained in the text.} \label{pressure}
\end{figure*}


\section{Basic formalisms for B-WDs}\label{sec1} 

The formalism of this paper is identical to the one used in Paper I, where stars are assumed to be approximately spherically symmetric.  It has already been shown that toroidally dominated stable magnetized WDs maintain their spherical symmetric shape to a very good approximation~\citep{Das2015a,Subramanian2015,Kalita2019}. Hence,  our restriction to spherically symmetric magnetized stars is not ad-hoc, since highly magnetized stars are already shown to be toroidally dominated (\citealt{Wickramasinghe2014,Quentin2018}). 
Moreover, the chosen strength of the magnetic field does not practically bring in any effects due to Landau quantization, see \citealt{Das2014a,Das2014b,Das2015a}. This in turn also assures minimal
or almost no effect on the symmetric structure of B-WDs. Note also that the magnetization has a negligible effect due to the magnetic field strength considered in this paper.  The contribution of magnetization to the pressure is at least an order of magnitude smaller than the magnetic field pressure,  i.e., $B^2/8\pi$. In fact, the minor effect of magnetization to pressure has been already explored in a few earlier works, see, e.g., ~\citealt{Ferrer2010, Sinha2013}.

The effective contributions from the matter $(\rho)$ and the magnetic field $(\rho_B)$ lead to the system density $(\tilde{\rho})$, given by
\begin{eqnarray}\label{1.3}
\widetilde{\rho} = \rho + \frac{B^2}{8\pi}.
\end{eqnarray}

The system pressure along the direction to the magnetic field is
represented as \emph{parallel pressure} and takes the form according
to magnetic field orientations as
\begin{eqnarray}\label{1.4}
 p_{\parallel} = \begin{cases}
 p_r - \frac{B^2}{8\pi}, \hspace{1cm} \textrm{for RO},\\
 p_t - \frac{B^2}{8\pi}, \hspace{1cm} \textrm{for TO}.
 \end{cases}
 \end{eqnarray}

Similarly, the system pressure that aligns perpendicular to the
magnetic fields is defined as \emph{transverse pressure} and, based on
the magnetic field orientations, takes the form 
\begin{eqnarray}\label{1.5}
 p_{\bot} = \begin{cases}
 p_t + \frac{B^2}{8\pi}, \hspace{1cm} \textrm{for RO},\\
 p_r + \frac{B^2}{8\pi}, \hspace{1cm} \textrm{for TO}.
 \end{cases}
 \end{eqnarray}
 
Based on the TOV equations, as shown in Paper I, and considering the magnetic field
orientations we obtain the essential magnetostatic stellar equations
which describe static, spherically symmetric B-WDs, given by
\begin{eqnarray}\label{1.8}
 \hspace{-4.5cm} \frac{\mathrm{d}m}{\mathrm{d}r} = 4\pi
  \left(\rho+\frac{B^2}{8\pi}\right) r^2,
\end{eqnarray}
\begin{eqnarray}\label{1.9}
 \hspace{-0.7cm} \begin{cases} {\frac {{\rm d}p_r}{{\rm d}r}}=\frac{-\left(\rho+p_{{r}}\right)\frac{4\pi{r}^{3}\left( p_{{r}}-   
  {\frac {{B}^{2}}{8\pi}}\right)+m}{r\left(r-2m\right)}+\frac{2}{r}\Delta}{\left[1-\frac{\rm d}{{\rm d}\rho}\left(\frac{B^2} 
  {8\pi}\right)\frac{{\rm d}\rho}{{\rm d}{p_r}}\right]}, \hspace{1.04cm}\textrm{for RO},\\ 
  {\frac {{\rm d}p_r}{{\rm d}r}}= \frac{-\left(\rho+p_{{r}}+\frac{B^2}{4\pi}\right)\frac{4\pi{r}^{3} \left( p_{{r}}+
  {\frac {{B}^{2}}{8\pi}} \right)+m}{r\left(r-2m\right)}+\frac{2}{r}\Delta}{\left[1+\frac{\rm d}{{\rm d}\rho}\left(\frac{B^2}
  {8\pi}\right)\frac{{\rm d}\rho}{{\rm d}{p_r}}\right]}, \hspace{0.4cm}\textrm{for TO}.
\end{cases} 
\end{eqnarray}
Here we describe the effective anisotropy of the stars with $\Delta$, which depends on the magnetic field orientations $p_t-p_r+\frac{B^2}{4\pi}$ in the case of RO and $p_t-p_r-\frac{B^2}{8\pi}$ for TO (Paper I). Note that the standard form of the TOV equation~\citep{Bowers1974,Herrera2013} for non-magnetized, anisotropic stars is retrieved by considering $B=0$.


\begin{figure}[!htpb]
\centering
\includegraphics[width=0.5\textwidth]{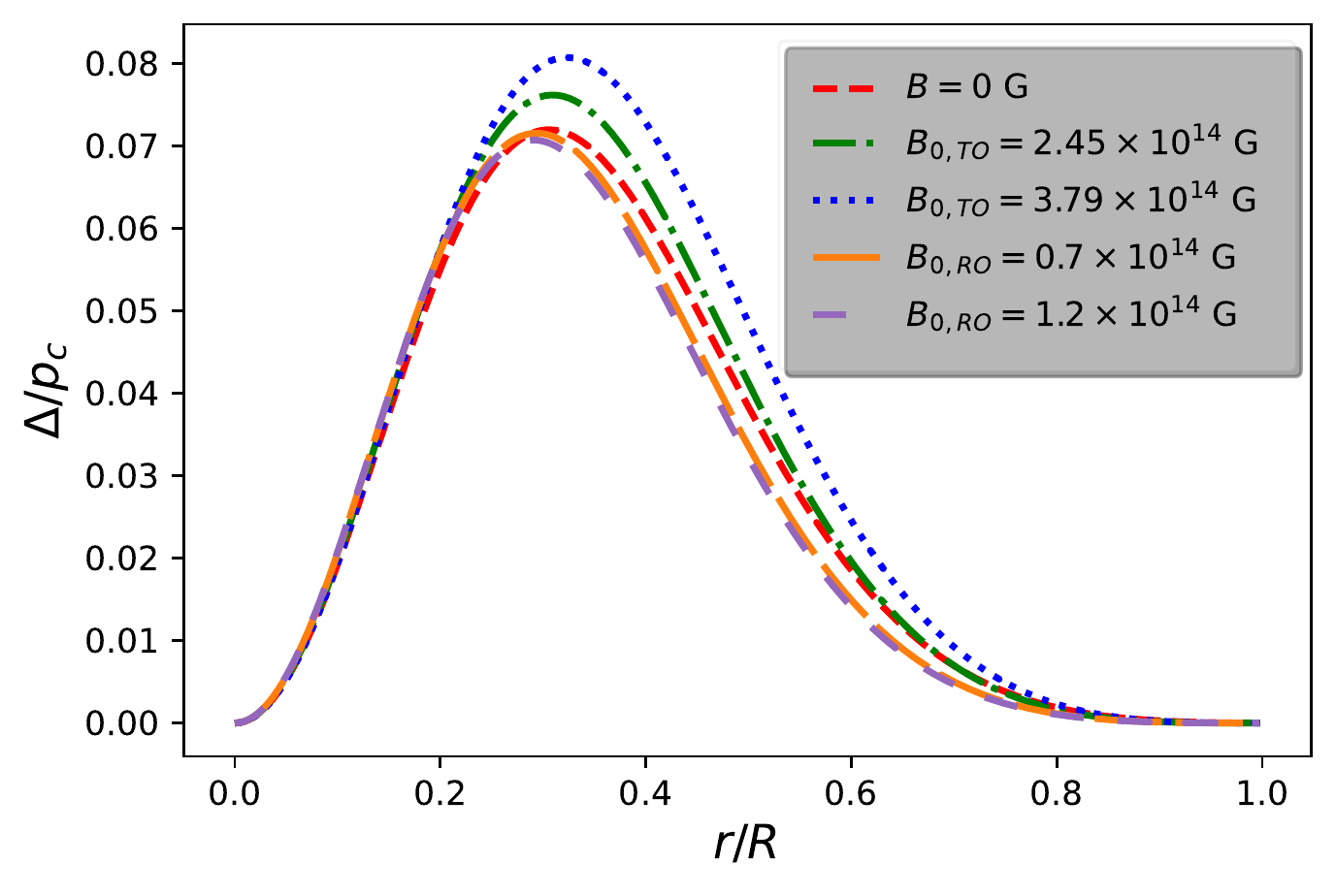}
\caption{Variation of anisotropy $(\Delta)$, normalized to central matter pressure $(p_c)$, as a function of radial distance $r/R$, for a magnetized $1.3\, M_\odot$ WD.} \label{aniso_WD1}
\end{figure}



\begin{figure}
\centering
\includegraphics[width=0.5\textwidth]{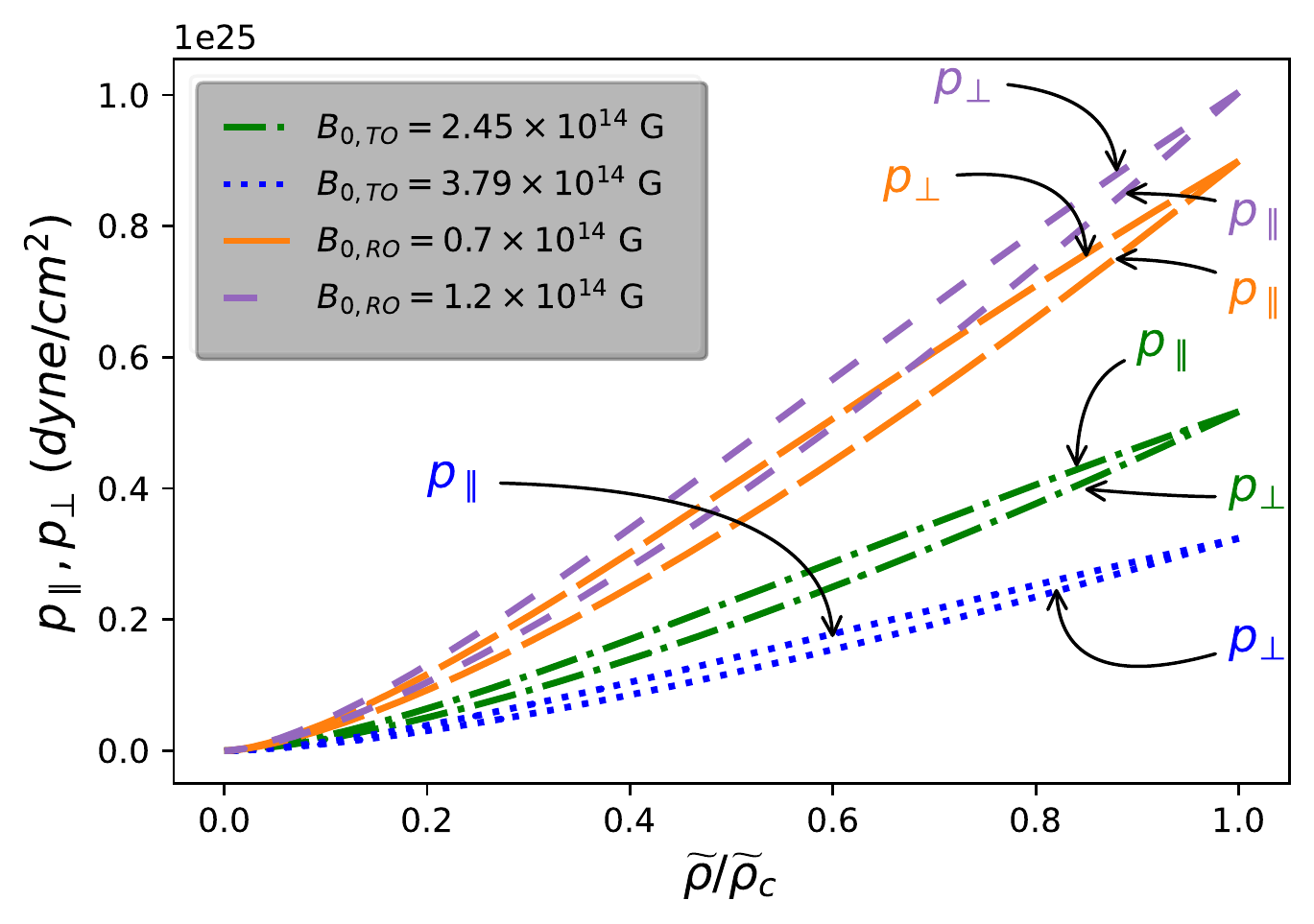}
\caption{ Variation of parallel pressure $(p_\parallel)$ and transverse  pressure $(p_\bot)$ with system density~$(\widetilde{\rho})$,  normalized to the central system density~$(\widetilde{\rho}_c)$, for  a magnetized $1.3\, M_\odot$ WD.
} \label{EOS}
\end{figure}


If one ignores local anisotropy due to the fluid, i.e.,  $p_t=p_r$, the right hand side of equation~\eqref{1.9}
diverges at $r=0$ as the central magnetic field $B_c$ is maximal at around center, which makes isotropic, highly magnetized WDs unstable. Since no exact theoretical form exists in the literature to deal with anisotropy, an approach that serves to avoid this instability has been proposed in Paper I. 

In Paper I, we already modeled the effective anisotropy that appeared in the TOV equations, which is the magnetized version of the
Bowers-Liang proposal (\citealt{Bowers1974}), given by
 \begin{eqnarray}\label{1.12}
 \Delta = \begin{cases}
 \kappa \frac{\left(\rho+p_r\right)\left(\rho+3\,p_r-\frac{B^2}{4\pi}\right)}{\left(1-\frac{2m}{r}\right)}r^2, \hspace{1.5cm} \textrm{for RO},\\
 \kappa \frac{\left(\rho+p_r+\frac{B^2}{4\pi}\right)\left(\rho+3\,p_r+\frac{B^2}{2\pi}\right)}{\left(1-\frac{2m}{r}\right)}r^2, \hspace{0.8cm} \textrm{for TO},
 \end{cases}
\end{eqnarray}
where $\kappa$ is a dimensionless constant which describes the strength of anisotropy within the stellar structure. $\kappa$ chosen in this work lies well within the permissible range of $\left[-\frac{2}{3},\frac{2}{3}\right]$~\citep{Silva2015}. Note importantly that $\kappa=0$ implies the anisotropy effects arisen due
to matter properties and magnetic field both vanish. However, the case of 
$B=0$ but $\kappa\neq 0$ implies only the anisotropy due to magnetic field vanishes.
In this work, we show that highly magnetized WD models which do  not account for the combined anisotropic effects of  the fluid and field are eliminated as they suffer an instability at the stellar center. Our phenomenological approach offers the best possible way to solve the (magneto-) hydrostatic equilibrium equations, which include anisotropic effects.


\begin{figure}[!htpb]
\centering
\includegraphics[width=0.5\textwidth]{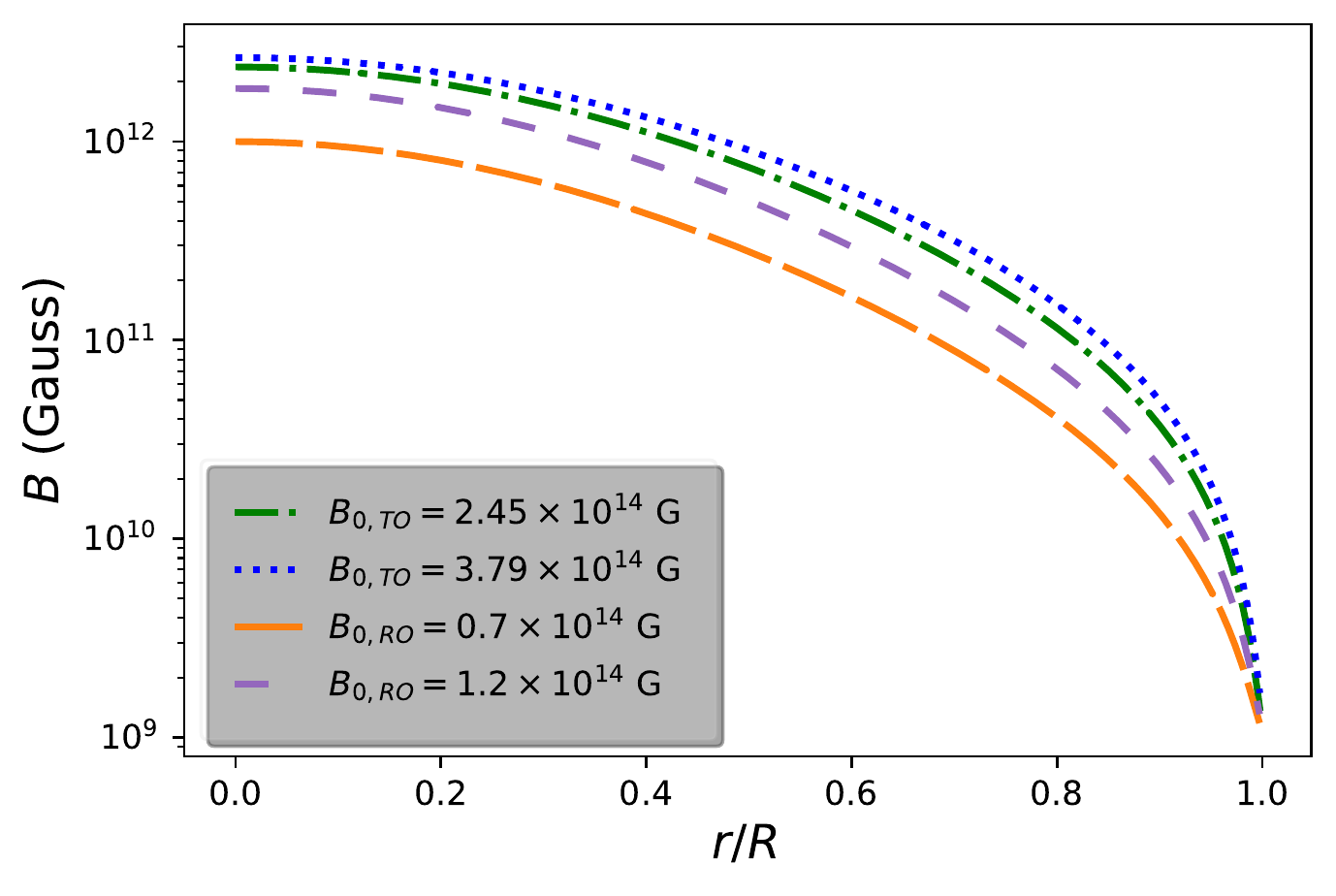}\hfill
\caption{Variation of magnetic field strength $B(\rho)$ with radial coordinate $r/R$, for a magnetized $1.3\, M_\odot$ WD.} \label{mag_field}
\end{figure}



\begin{figure*} 
\centering
\includegraphics[width=0.33\textwidth]{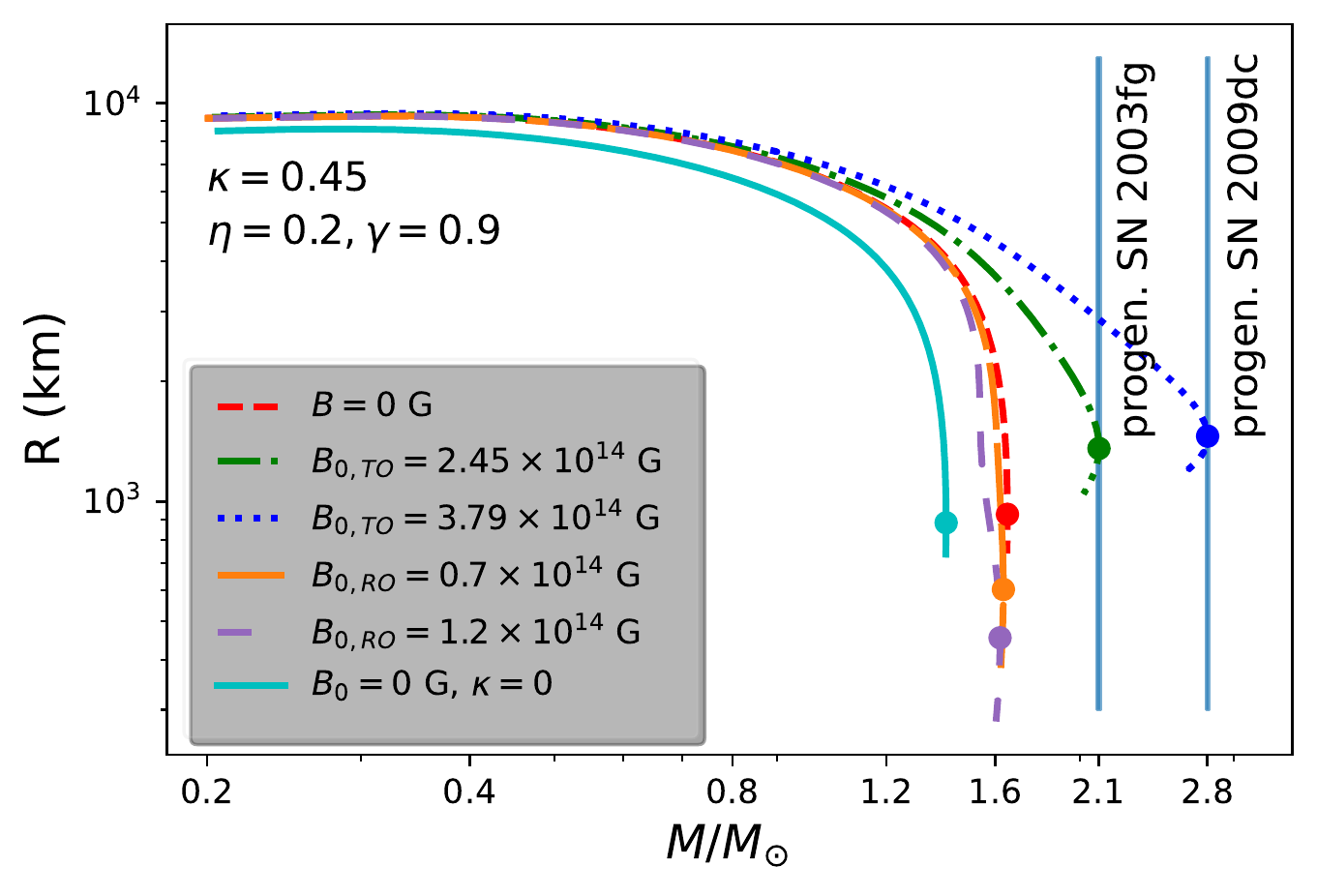}\hfill
\includegraphics[width=0.33\textwidth]{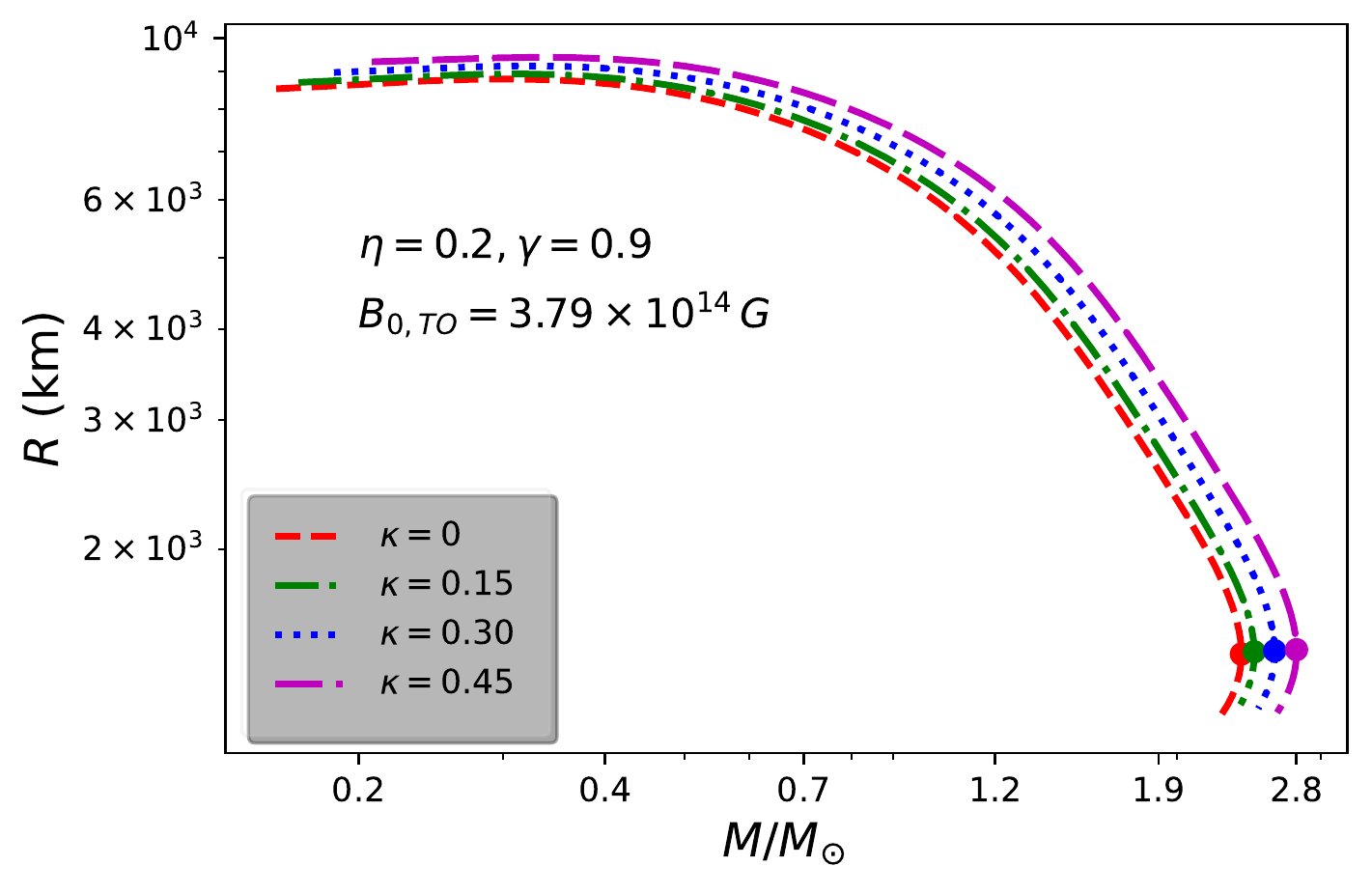}\hfill
\includegraphics[width=0.33\textwidth]{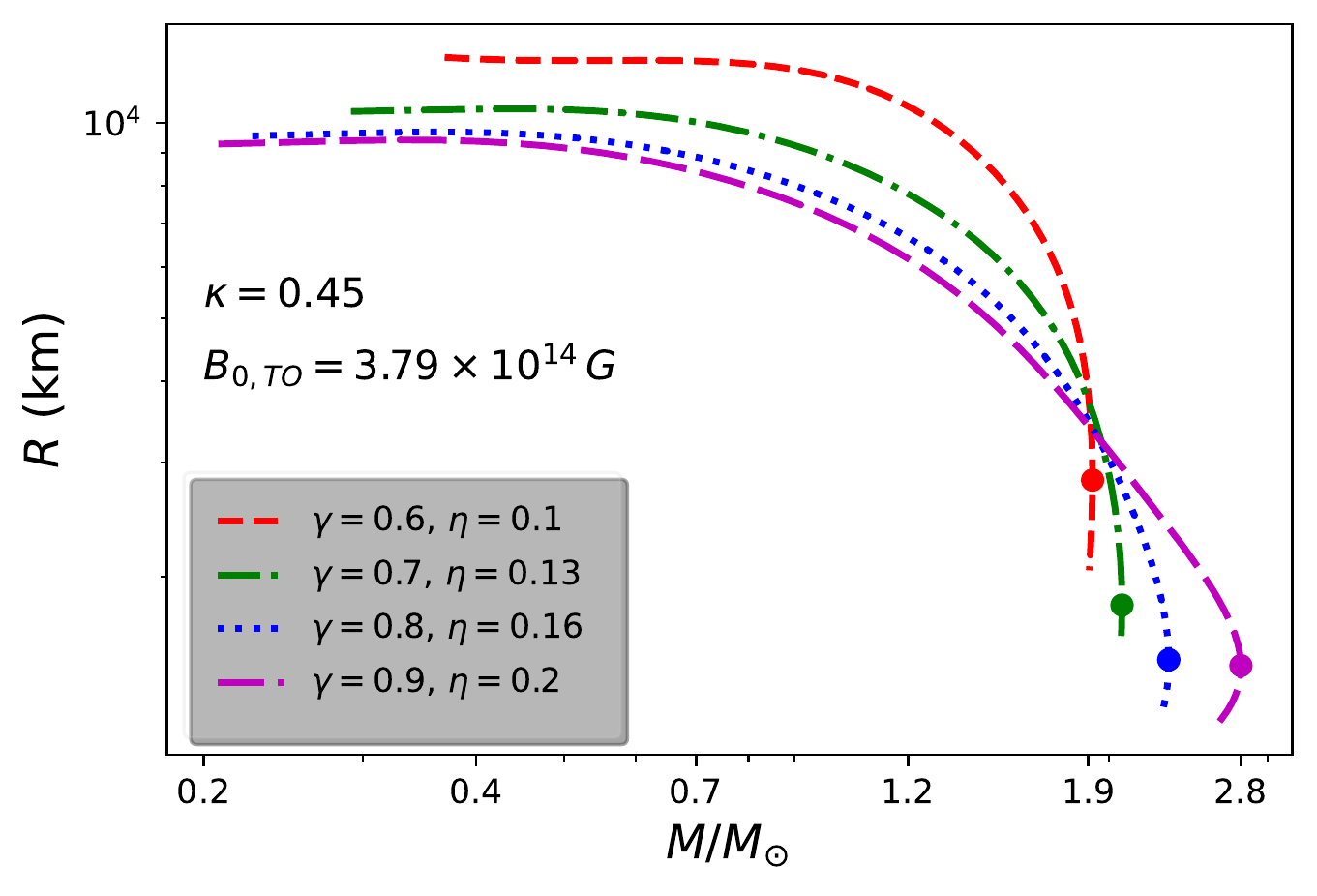}
\caption{Stellar radius $(R)$ as a function of
	gravitational mass $(M/M_\odot)$ for different (a) $B_0$ (left panel),  (b) $\kappa$ (middle panel), and (c) $\eta$ and $\gamma$ (right panel). Solid circles represent the stars with the
    maximum-possible masses.}\label{MR}
\end{figure*}



\begin{figure*}
\centering
\includegraphics[width=0.33\textwidth]{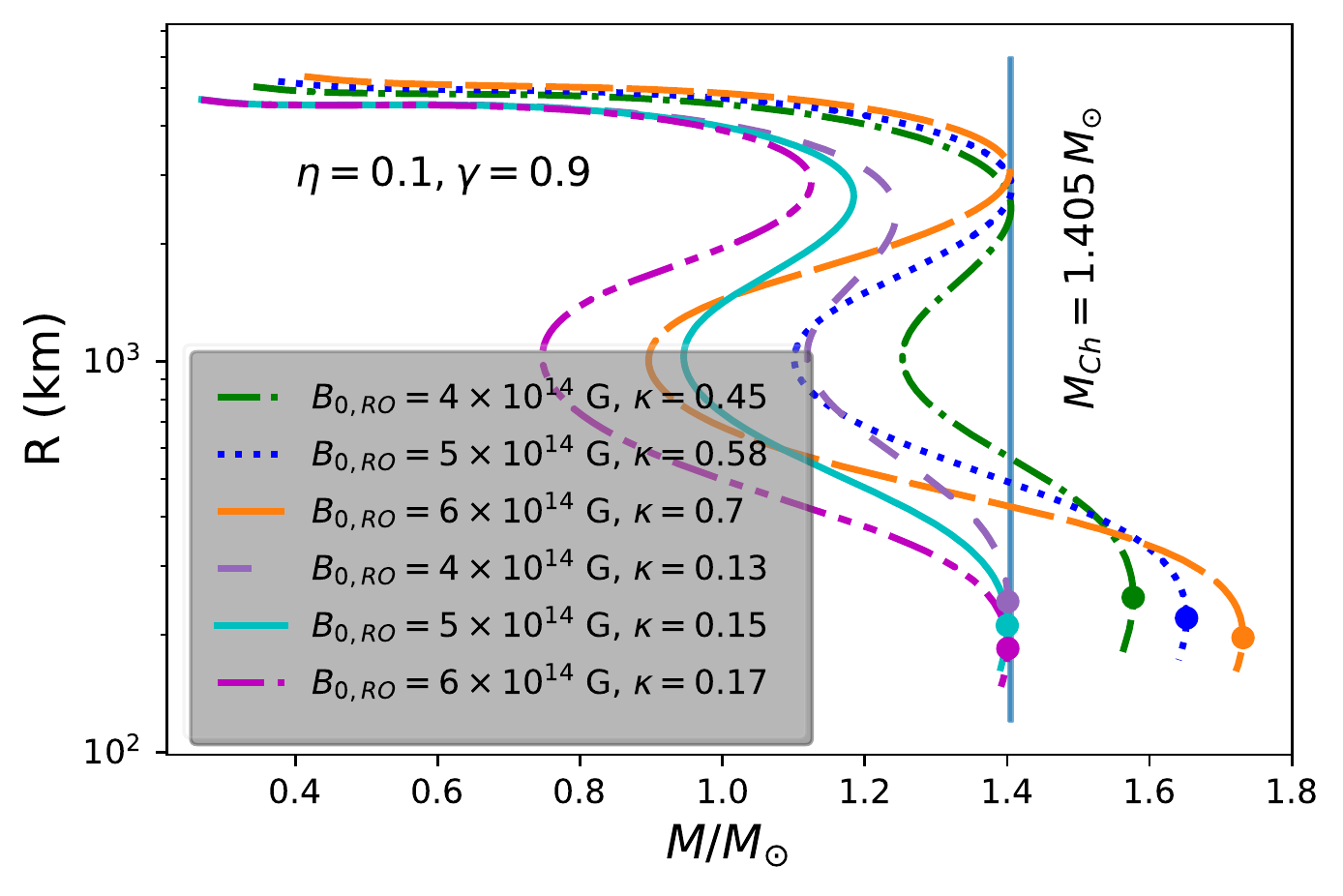}\hfill
\includegraphics[width=0.33\textwidth]{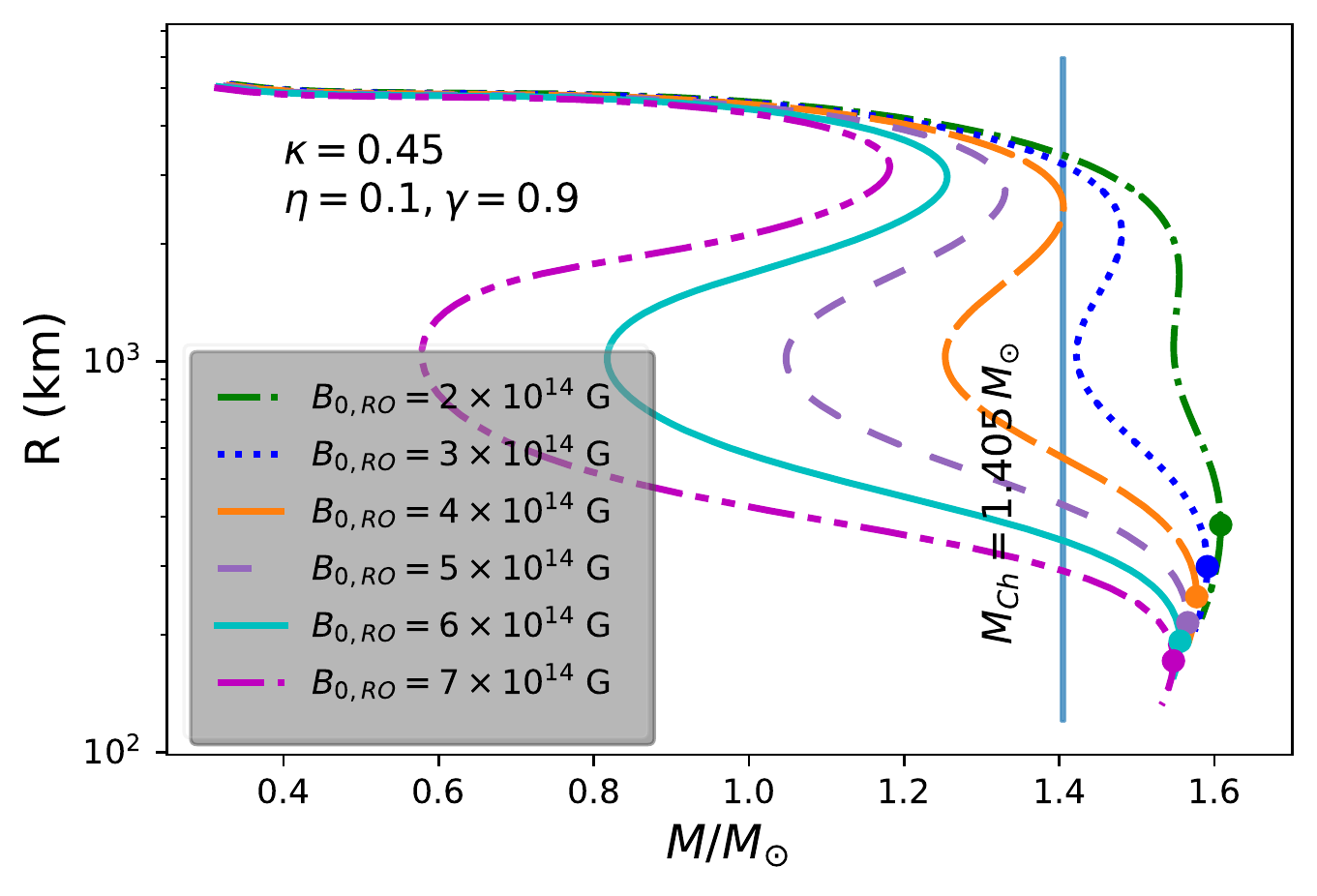}\hfill
\includegraphics[width=0.33\textwidth]{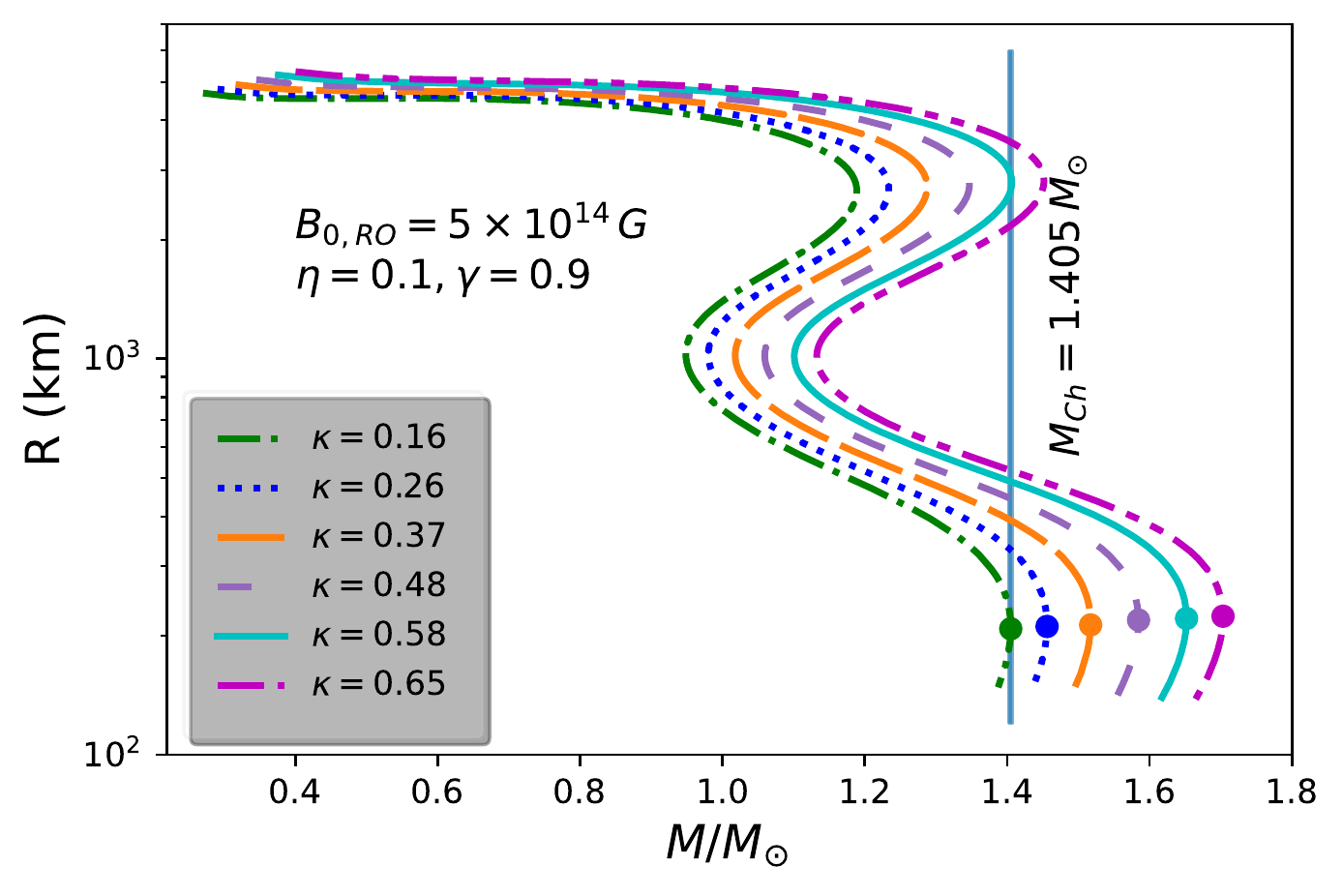}
\caption{ Stellar radius $(R)$ as a function on gravitational mass $(M/M_\odot)$  for varying (a) $B_{0, \textrm{RO}}$ and $\kappa$ (left panel), (b)  $B_{0, \textrm{RO}}$ (middle panel), and (c)  $\kappa$ (right panel). Solid circles represent the stars with the  maximum-possible masses. } \label{MR_test}
\end{figure*}


To solve the (magneto-)hydrostatic stellar structure equations~\eqref{1.8} and~\eqref{1.9} from the stellar center to surface, it is required to supply an equation of state (EoS), which connects $\rho$ with $p_r$, along with the functional form of $\Delta$. In this work we consider an EoS proposed by Chandrasekhar to explain WDs supported by electron degeneracy pressure, given by
\begin{eqnarray}\label{1.13}
 \hspace*{-1cm} p_r = \frac{\pi m^4_e c^5}{3 h^3}\left[x\left(2 x^2
   -3\right)\sqrt{x^2+1}+3 \; \mathrm{sinh}^{-1}x\right],\nonumber
 \\\label{1.14} \rho = \frac{8\pi\mu_e m_H \left(m_e c\right)^3}{3
   h^3} x^3,
\end{eqnarray}
where $m_e$ is the mass of an electron, ${m_H}$ is the mass of a hydrogen atom, $h$ is Planck's constant, $\mu_e$ is the mean molecular weight per electron, and $x=p_F/{m_e} c$ with $p_F$ the Fermi
momentum. For the description of C-O WDs in this work we set $\mu_e=2$. To solve the coupled non-linear differential equations~\eqref{1.8} and~\eqref{1.9} we consider boundary conditions at (i) the stellar center $m(r)|_{r=0}=0$ and $\rho(r)|_{r=0}=\rho_c$ and (ii) the stellar surface $\rho(r)|_{r=R}=0$, which ensures the essential junction condition proposed by \cite{OBrien1952} and \cite{Robson1972}, i.e., $p_r=0$. 
To describe the exterior spacetime we consider the Schwarzschild metric.

It is generally known that the central magnetic field strength $(B_c)$ is several orders of magnitude higher than $B_s$~\citep{Fujisawa2012,Das2014b,Subramanian2015}. This phenomenon is due to the presence of a  fossil  field in the progenitor star, which is supposed to be stronger at the stellar core than at the surface, and the dynamo effect which ensures a strong field at the center~\citep{Potter2010}. This feature is taken into account by adopting the density-dependent magnetic field profile of Paper I, which decreases monotonically inside B-WDs, from its maximum finite value at the center to its minimum value at the surface. This appropriately mimics the spatial dependence of the magnetic field strength. The chosen density-dependent magnetic field profile, which was originally proposed by~\cite{Bandyopadhyay1997,Bandyopadhyay1998a} and applied later to WDs by~\cite{Das2014b,Bhattacharya2018}, is given by
\begin{eqnarray}\label{1.16}
B(\rho)= B_s + B_0\left[1-\exp\left\lbrace-\eta\left(\frac{\rho}{\rho_0}\right)^\gamma\right\rbrace\right],
\end{eqnarray}
where the dimensionless parameters $\eta$ and $\gamma$
control how  $B(\rho)$ decreases from the center to the
surface of a magnetized star. Here, following
\cite{Bhattacharya2018} we chose $\rho_0=10^9\,\mathrm{g/cm^{3}}$,
$\gamma=0.9$ and $\eta=0.1$.  We also assume that $B_s=10^9$~G, which is consistent with the observations made by the Sloan Digital Sky Survey~\citep{Schmidt2003,Kepler2015,Ferrario2020}. Nevertheless, our results are not sensitive
to $B_s\lesssim 10^9$ G, as shown previously by, e.g., \cite{Das2015a,Gupta2020}. 

However, ~\cite{Dexheimer2017} later proposed that density-dependent magnetic field strength profile should be polynomial instead of exponential which we may explore in future works. The validity of Maxwell's equation for this choice of magnetic field profile with respect to the chosen field orientations was discussed in detail in Paper I and will thus not be repeated here.

\section{Results and Discussions}\label{sec2}

Our study reveals that the orientations of magnetic fields (such as RO and TO) significantly affect the structural and interior properties of B-WDs. To feature the combined effects of anisotropy, magnetic fields and their orientations to the 
said structural and interior properties, we take a WD candidate with a mass of ~$1.3\,M_{\odot}$ and surface magnetic field $10^9~$G in Figs. \ref{pressure}, \ref{aniso_WD1}, \ref{EOS}, \ref{mag_field}. Since the effect of the magnetic field on the stellar mass is not same for  RO and TO fields,  we choose different $B_{0}$ for the TO and RO fields. TO and RO in the subscript of $B_0$ denote the associated magnetic field orientations. Moreover, the values of (constant) free parameters have been chosen such that the progenitors of observed peculiar over-luminous SNeIa can be explained.


\begin{figure}[!htpb]
\centering
\includegraphics[width=0.5\textwidth]{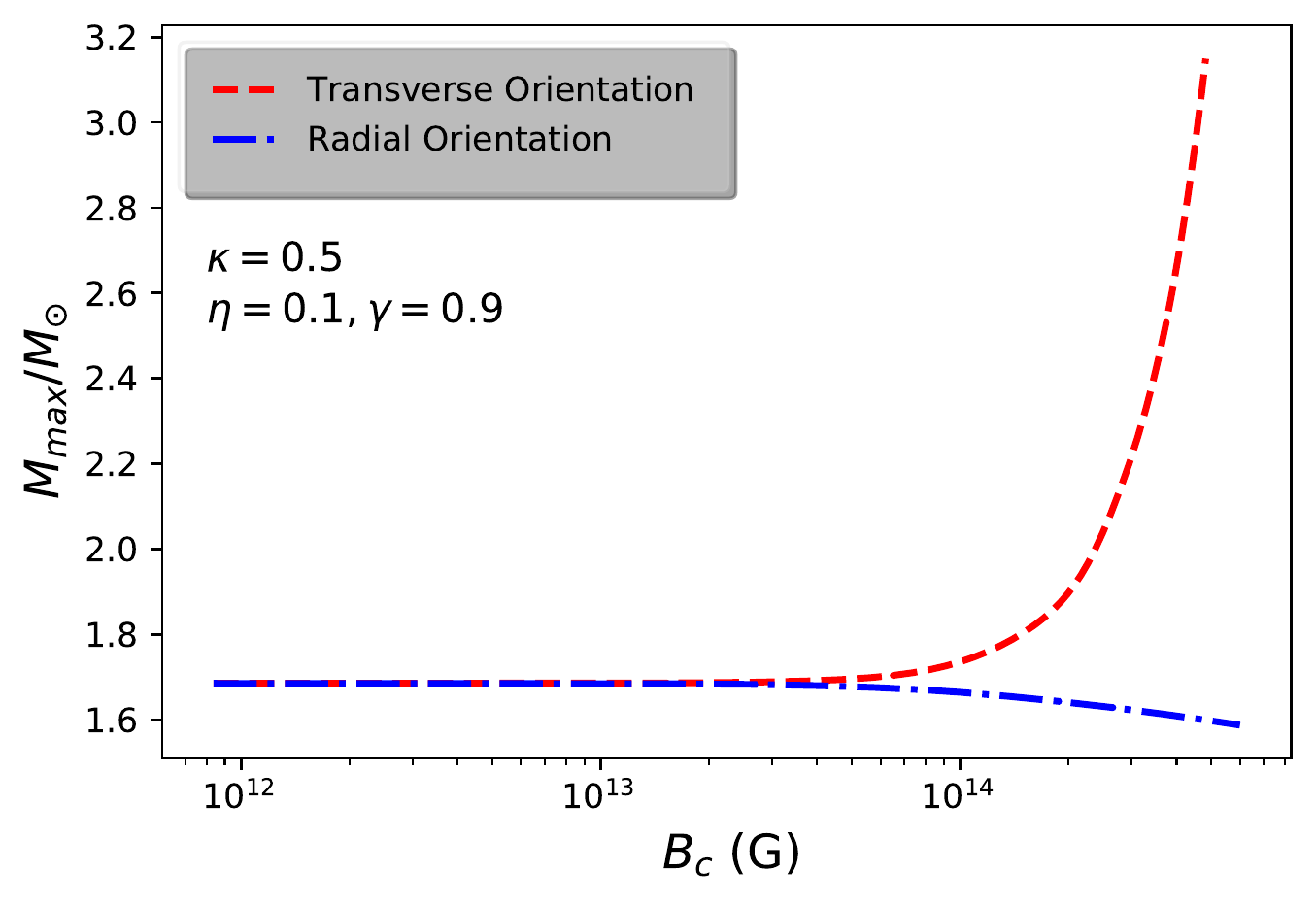}
\caption{Variation of stellar mass $(M/M_\odot)$ with central magnetic field $(B_c)$.} \label{massmag}
\end{figure}



\begin{figure}[!htpb]
\centering
\includegraphics[width=0.5\textwidth]{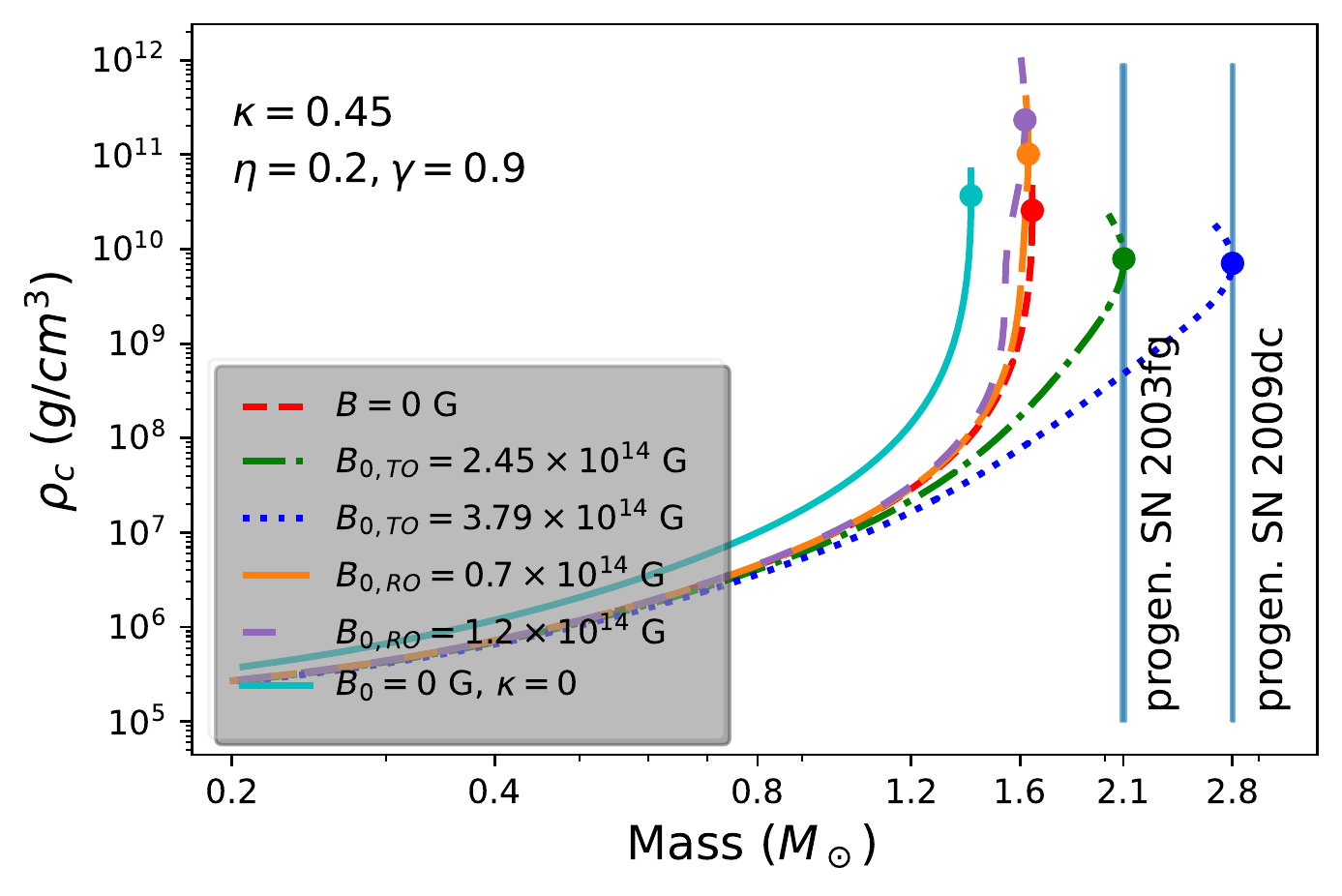}
\caption{Variation of central mass density $(\rho_c)$ with stellar mass $(M/M_\odot)$. Solid circles represent stars with maximum possible masses.} \label{masscden}
\end{figure}



\begin{figure}[!htpb]
\centering
\includegraphics[width=0.5\textwidth]{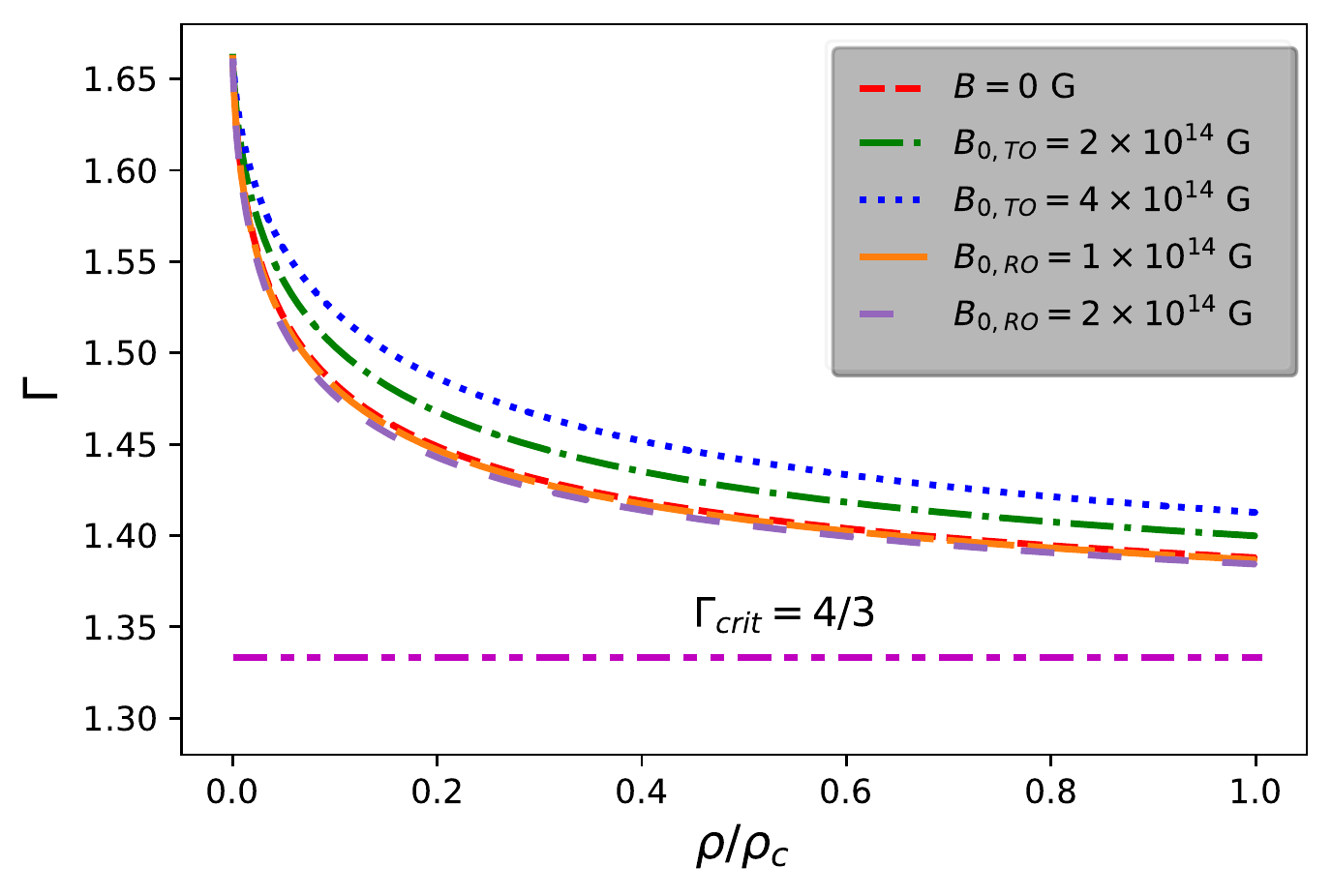}
\caption{Variation of adiabatic index ($\Gamma$) with radial coordinate ($r/R$) of a $1.3\, M_\odot$ magnetized WD model.} \label{adia}
\end{figure}


The profiles for matter density $\rho$ with respect to normalized radial coordinate $r/R$ are shown in the left panel of Fig.~\ref{pressure}, where $R$ is the radius of WDs. The profiles of the radial pressure $p_r$ and tangential pressure $p_t$ of the matter are shown in the middle and right panels of Fig.~\ref{pressure}, respectively. Figure~\ref{pressure} shows that $\rho$, $p_r$ and $p_t$ have maximum finite values at the stellar center which decrease gradually to reach their minimum values at the surface, which also ensures physical regularity of the proposed model. It is also evident from Fig.~\ref{pressure} that this proposed stellar model is free from a gravitational singularity (or spacetime singularity). In Fig.~\ref{aniso_WD1} we show the profiles of anisotropy ($\Delta$) in WDs for TO and RO  magnetic fields. One sees that the maximal anisotropic stress inside a TO WD  increases  with $B_0$. This is in contrast to RO fields, in which case the maximal anisotropic stress decreases for increasing  $B_0$. Importantly, the anisotropy is zero at the center of B-WDs, which ensures consistency of the equilibrium of forces at every interior stellar location, from the center to the surface of the stellar structure. One can see that for TO fields, in the case of $B_0=3.79\times {10}^{14}$~G, the maximum effective anisotropy is $\sim$92\% lower in magnitude compared to the matter central pressure $p_c$ which is minuscule to drive the stellar configuration towards non-spherical symmetry.

In Fig.~\ref{EOS} we show the combined effects of the magnetic field and its orientation on the parallel pressure $p_\parallel$ and transverse pressure $p_\bot$. One sees that as $B_0$ increases for TO fields, the slope of the system pressure profile decreases due to the increase of the size of the star and the decrease of the outward hydrodynamic force ($F_h$). In contrast, the system pressure profiles stiffen for increasing $B_0$ in the RO case as the size of the star decreases and $F_h$ gradually increases. Note that at the center and surface of the stars, $p_\parallel$ and $p_\bot$ have the same value, reflecting the
consistency of the TOV equation of highly magnetized B-WDs in the present treatment. This was overlooked by almost all the researchers before Paper I. On the other hand, it has already been shown that for toroidal fields B-WDs approximately maintain their spherically symmetric configuration~\citep{Das2015a,Subramanian2015,Kalita2019}. We show the density-dependent magnetic field profiles for different TO and RO fields in Fig.~\ref{mag_field} which, as expected, are maximal at the center and decrease gradually within B-WDs to reach their minimum values at the surface.


\begin{figure}[!htpb]
\centering
\includegraphics[width=0.5\textwidth]{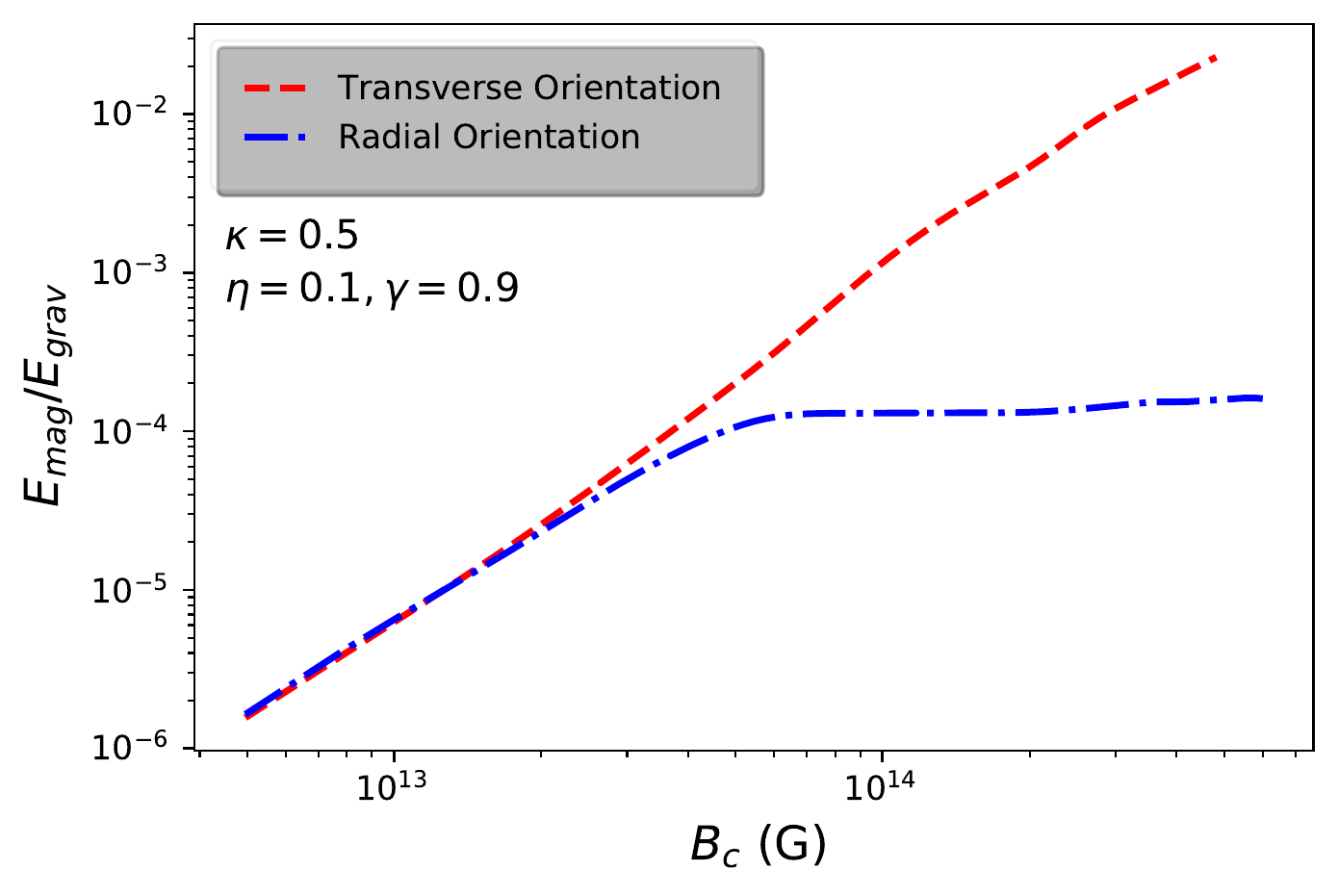}
\caption{Variation of the ratio of the magnetic energy $(E_{\rm mag})$ to  gravitational energy $(E_{\rm grav})$ with central magnetic field $(B_c)$.} \label{MEGE}
\end{figure}


In Fig.~\ref{MR}, we show the mass--radius relations of B-WDs for different $B_0$, $\kappa$, and $\gamma$. Our study shows that for TO fields with $B_0=3.79\times{10}^{14}$~G, a maximum mass B-WD of $2.8\, M_\odot$ is obtained whose radius is 1457.67~km.  For a RO field with $B_0=1.2\times {10}^{14}$~G, the maximum mass drops to $1.62\, M_\odot$ and the radius of the WD is
454.67~km, as shown in the left panel of Fig.~\ref{MR}. We find for $B_{0,\mathrm{TO}}=3.79\times{10}^{14}$~G that the maximum mass and the corresponding radius of B-WDs increase by $\sim$70\% and $\sim$57\%, respectively, compared to the non-magnetized but anisotropic case. For $B_{0,\mathrm{RO}}=1.2\times{10}^{14}$~G, the maximum mass and the corresponding radius decrease by $\sim$2\% and $\sim$52\%, respectively, compared to the values of non-magnetized but anisotropic WDs. Note that without considering the magnetic field and incorporating the effects of local anisotropy due to the fluid, it is possible to push the maximum mass of WDs beyond the CML. For example, by considering $\kappa=2/3$, we obtain a maximum mass for a non-magnetized but anisotropic WD of $1.81~M_{\odot}$. The corresponding radius is 956.08~km. These values are $\sim$29\% and $\sim$8\%, respectively, higher than the respective values of WDs at the CML. One sees an almost similar trend in the middle panel of Fig.~\ref{MR} which shows that as the free parameter $\kappa$ increases for a fixed $B_{0,\mathrm{TO}}=3.79\times {10}^{14}$~G, the maximum masses of WDs increase. Further, with increasing $\eta$ and $\gamma$ for the TO field case, the mass of anisotropic B-WDs changes significantly, as can be seen in the right panel of Fig.~\ref{MR}. It shows that the maximum mass of a B-WD with $\gamma=0.9$ and $\eta=0.2$ increases by $\sim$46\% compared to $\gamma=0.6$ and $\eta=0.1$.


\begin{table*}
  \centering
	\caption{Physical parameters of WDs with $B_s=10^9$ G, $\kappa = 0.45$,
      $\eta=0.2$ and $\gamma=0.9$ for different $B_0$.} \label{Table
      1}
\begin{tabular}{ ccccccccccccccccccccccccccc}
\hline\hline

Orientation & & & Corresponding & Central & Central & Central & \\ of
magnetic & $B_0$ & Maximum & predicted & magnetic field & density &
pressure & $\frac{E_{\rm mag}}{E_{\rm grav}}$\\ field & (G) & mass
$(M_\odot)$ & radius (km) & $B_c$ (G) &
$\widetilde{\rho}_c~({\rm g/cm}^3)$ & $\widetilde{p}_c~({\rm dyne/cm}^2)$ &
& \vspace{0.2cm} \\ \hline
\multirow{2}{*}{Transverse Orientation}              & $3.79 \times {10}^{14}$ & $2.80$ & $1457.67$ & $2.606\times {10}^{14}$ & $7.079\times {10}^{9}$  & $2.398\times {10}^{28}$   & $1.48\times {10}^{-2}$ \\ 
\ 
                                                     & $2.45 \times {10}^{14}$ & $2.10$ & $1357.96$ & $1.772\times {10}^{14}$ & $7.903\times {10}^{9}$  & $3.389\times {10}^{28}$   & $6.70\times {10}^{-3}$ \\ 

No magnetic Field                                    & -                       & $1.65$ & $927.90$  & - & $2.569\times {10}^{10}$ & $9.865\times {10}^{28}$ & - \\ 

\multirow{2}{*}{Radial Orientation}                  & $0.7 \times {10}^{14}$  & $1.63$ & $601.27$  & $0.7\times {10}^{14}$ & $1.015\times {10}^{11}$  & $1.565\times {10}^{30}$   & $6.04\times {10}^{-5}$ \\ 

                                                     & $1.2 \times {10}^{14}$  & $1.62$ & $454.67$  & $1.2\times {10}^{14}$ & $2.328\times {10}^{11}$  & $5.344\times {10}^{30}$   & $7.75\times {10}^{-5}$\\

\hline\hline

\end{tabular}  
  \end{table*}



\begin{table*}
  \centering
	\caption{Physical parameters of WDs with $B_0 = 3.79\times{10}^{14}$ G, $B_s=10^9$ G, $\eta=0.2$ and $\gamma=0.9$ for different
      $\kappa$. } \label{Table 2}
\begin{tabular}{ ccccccccccccccccccccccccccc}
\hline\hline

    &     & Corresponding & Central & Central & Central &    \\
 $\kappa$ & Maximum & predicted & magnetic field & density & pressure & $\frac{E_{\rm mag}}{E_{\rm grav}}$\\
       &  mass $(M_\odot)$ & radius~(km) & $B_c$~(G) & $\widetilde{\rho}_c~({\rm g/cm}^3)$ & $\widetilde{p}_c~(\rm{dyne/cm}^2)$ &  \\ 
\hline 
 $0$    & $2.36$ & $1443.23$ & $2.812\times {10}^{14}$ & $8.380\times {10}^{9}$ & $2.867\times {10}^{28}$ & $1.66\times {10}^{-2}$ \\ 
\
 $0.15$ & $2.49$ & $1447.92$ & $2.744\times {10}^{14}$ & $7.922\times {10}^{9}$ & $2.852\times {10}^{28}$ & $1.63\times {10}^{-2}$ \\ 
\
 $0.30$ & $2.63$ & $1452.73$ & $2.676\times {10}^{14}$ & $7.488\times {10}^{9}$ & $2.837\times {10}^{28}$ & $1.55\times {10}^{-2}$ \\ 
\
 $0.45$ & $2.80$ & $1457.67$ & $2.606\times {10}^{14}$ & $7.079\times {10}^{9}$ & $2.823\times {10}^{28}$ & $1.48\times {10}^{-2}$\\

\hline\hline

\end{tabular}  
  \end{table*}



\begin{table*}[htbp!]
  \centering
	\caption{Physical parameters of WDs with $B_0 = 3.79\times{10}^{14}$ G, $B_s=10^9$ G, and $\kappa = 0.45$ for different $\eta$ and $\gamma$.} \label{Table 3}
\begin{tabular}{ ccccccccccccccccccccccccccc}
\hline\hline

        &        & Corresponding & Central & Central & Central &    \\
 $\gamma$ and $\eta$ & Maximum & predicted & magnetic field & density & pressure & $\frac{E_{\rm mag}}{E_{\rm grav}}$\\
      &  mass $(M_\odot)$ & radius~(km) & $B_c$ (G) & $\widetilde{\rho}_c~({\rm g/cm}^3)$ & $\widetilde{p}_c~({\rm dyne/cm}^2)$ &  \\ 
\hline 
 $\gamma=0.6,\,\eta=0.1$    & $1.92$ & $2815.92$ & $6.077\times {10}^{13}$ & $8.610\times {10}^{8}$ & $1.958\times {10}^{27}$ & $1.62\times {10}^{-1}$ \\ 
\
 $\gamma=0.7,\,\eta=0.13$ & $2.07$ & $1807.05$ & $1.364\times {10}^{14}$ & $3.603\times {10}^{9}$ & $5.820\times {10}^{27}$ & $3.79\times {10}^{-2}$ \\ 
\
 $\gamma=0.8,\,\eta=0.16$ & $2.33$ & $1488.71$ & $2.528\times {10}^{14}$ & $6.646\times {10}^{9}$ & $1.815\times {10}^{28}$ & $1.90\times {10}^{-2}$ \\ 
\
 $\gamma=0.9,\,\eta=0.2$ & $2.80$ & $1457.67$ & $2.606\times {10}^{14}$ & $7.079\times {10}^{9}$ & $2.823\times {10}^{28}$ & $1.48\times {10}^{-2}$\\

\hline\hline

\end{tabular}  
  \end{table*}


Although the inclusion of rotation can explain SCPWD up to $1.7\, M_\odot$, the combined effects of the magnetic field and rotation on WDs can push the maximum mass to much high values. However, all these models can not simultaneously explain sub-Chandrasekhar progenitor WDs. In fact, except the model proposed by~\cite{Das2015b,Kalita2018}, in the realm of modified gravity, no other model can simultaneously explain both the super- and sub-Chandrasekhar WDs till this date. Importantly, in this work for RO fields we are able to explain sub-, standard- and super-Chandrasekhar B-WDs by making appropriate choices of $B_{0,\mathrm{RO}}$ and $\kappa$ (see Fig.~\ref{MR_test}). By changing both $B_{0,\mathrm{RO}}$ and $\kappa$, as shown in the left panel of Fig.~\ref{MR_test}, we successfully explain both (i) the sub- and standard-Chandrasekhar progenitor B-WDs and (ii) the standard and super-Chandrasekhar progenitor B-WDs,  by a single mass--radius curve for the respective cases. On the other hand, through changes of only $B_{0,\mathrm{RO}}$ or $\kappa$,  sub-, standard- and super-Chandrasekhar B-WDs line up in a series of mass--radius curves, as shown in the middle and right panels of Fig.~\ref{MR_test}. This leads to a complete explanation of under-, regular- and over-luminous SNeIa in a single theory.

From Fig.~\ref{massmag} it is evident that the effects of the magnetic field strength on WDs are not the same for RO and TO magnetic fields, which is the reason for the choice of different $B_0$ for different field orientations. For example,  for a reference value of $B_c=4\times {10}^{14}$~G and the particular choices $\kappa=0.5$, $\eta=0.1$ and $\gamma=0.9$, the asymmetry between the maximum masses obtained for RO and TO fields is 65.4\% which indicates how significant the influence of the magnetic field orientations is on WDs. Interestingly, with the increase of $B_{0,\mathrm{TO}}$ the maximum mass, $M_{max}$, of B-WDs increases rapidly, whereas the decrease of $M_{max}$ for increasing $B_{0,\mathrm{RO}}$ turns out to be not very significant, as shown in Fig. \ref{massmag}.

To present a stable spherically symmetric stellar model, the model must be consistent with the inequality $\mathrm{d}M/\mathrm{d}\rho_\mathrm{c}>0$~\citep{Harrison1965} up to its maximum mass for a mass--radius relation. This stability criterion is well satisfied in this work up to the maximum mass star, as shown in Fig.~\ref{masscden}. To ensure dynamical stability of spherically symmetric compact stars against an infinitesimal radial adiabatic oscillation,~\cite{Chandrasekhar1964a,Chandrasekhar1964b} introduced a classic technique based on the adiabatic index ($\Gamma$) of the system. This result was later revisited and reinstated by~\cite{Heintzmann1975} who showed that even anisotropic spherically symmetric  compact stars should maintain $\Gamma>4/3$ at all the interior points of the stars to ensure dynamical stability of the stellar structure. We show in Fig.~\ref{adia} that our model is consistent with $\Gamma>4/3$ at all the interior points of B-WDs. However,~\cite{Chandrasekhar1953} showed that although it is necessary to be consistent with $\Gamma>4/3$, which is not sufficient to ensure dynamical stability for the spherically symmetric magnetized compact stars as the presence of sufficiently strong magnetic field may lead to instability within the stellar structure. They showed that to achieve dynamical stability a spherically symmetric magnetized stellar structure further needs to be consistent with the stability criteria $\mid E_{\rm grav} \mid>E_{\rm mag}$, where $\mid E_{\rm grav} \mid$ denotes the gravitational potential energy and $E_{\rm mag}$ represents magnetic energy. Through Fig.~\ref{MEGE} we show that $\mid E_{\rm grav} \mid$ significantly overpowers $E_{\rm mag}$, which confirms dynamical stability of the proposed B-WD models.

\begin{figure}
\centering
\includegraphics[width=0.5\textwidth]{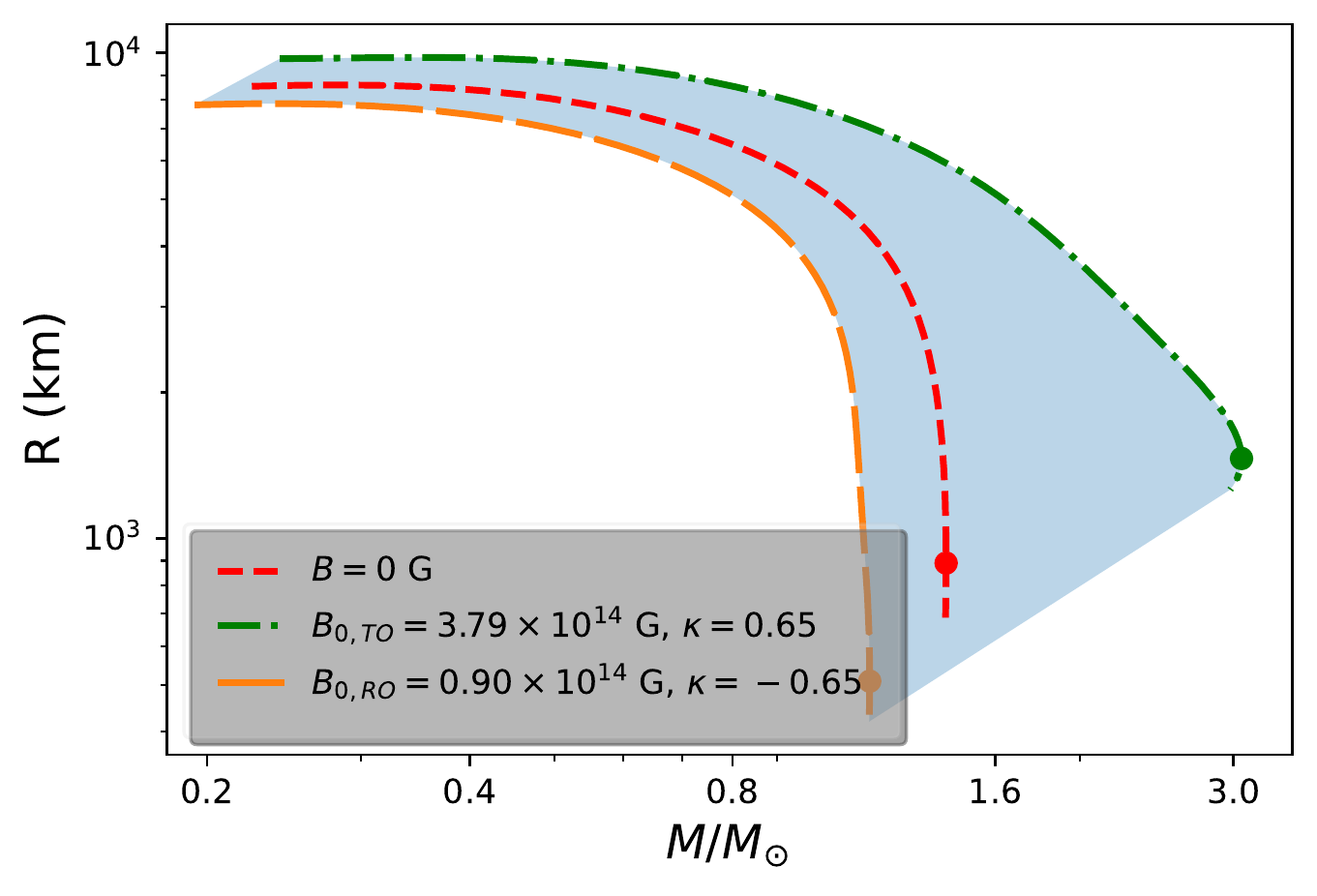}
\caption{Variation of radius $(R)$ with stellar mass $(M/M_\odot)$. Solid circles represent maximum possible stars.} \label{negk}
\end{figure}

We show in Tables~\ref{Table 1}-\ref{Table 3} how the combined effects of anisotropy, magnetic field strength and its orientations have noteworthy influence on the different physical parameters of B-WDs. Table~\ref{Table 1} shows the changes of $M_{\rm max}$, $R$,  $B_c$, system central density ($\tilde{\rho}_c$), system central pressure ($\tilde{p_c}$) and $\mid E_{\rm grav} \mid / E_{\rm mag}$ for varying $B_0$, where we have chosen $\kappa=0.45$, $\eta=0.2$ and $\gamma=0.9$. In Table~\ref{Table 2}, we show the change of the said physical parameters for varying $\kappa$. Table~\ref{Table 3} shows how the physical properties of B-WDs depend on parameters such as $\eta$ and $\gamma$. Throughout our work we have considered only positive  $\kappa$. One however can see that our model with negative $\kappa$ (see \citealt{Silva2015}, for exploration of negative $\kappa$ in neutron stars) for RO fields can suitably explain sub-Chandrasekhar limiting mass WDs. In Fig.~\ref{negk} we show the mass--radius relations for the lower and upper bounds of $\kappa$. Importantly, we can explain any WD which lies within the blue shaded area shown in Fig.~\ref{negk}.


\section{Conclusion}\label{sec3}
We have analysed the combined effects of anisotropy, magnetic field strengths and field orientations on B-WDs. As already pointed out by~\cite{Chowdhury2019},  magnetic fields are one of the key reasons for anisotropy within WD stars. These authors proposed a spherically symmetric model for  anisotropic WDs without  considering the presence of a magnetic field. In a separate study~\cite{Chu2014}, as well as Paper I, showed that RO and TO fields could reduce and enhance, respectively, the maximum mass of strange quark stars. To the best of our knowledge, we study, for the first time in the literature, the properties of spherically symmetric anisotropic B-WDs. As demonstrated, the effective anisotropy, magnetic field strength and its orientations have a significant influence on the properties of strongly magnetized WDs. This work manifests many fold important outcomes as follows. 

(i) Through this work, we show that to maintain hydrodynamic stability at the stellar core for B-WDs, it is important to consider the combined effects of anisotropy due to the fluid and the magnetic field.

(ii) By choosing an appropriate set of constant free parameters, such as $B_0$ and $\kappa$, it is possible to explain highly massive progenitors of peculiar over-luminous SNeIa. This immediately questions the idea that the $1.4\, M_\odot$ white dwarf is related to ``standard candle", that is used as an important tool to verify the contemporary idea of accelerated expansion of the universe.

(iii) We are able to explain the possible combinations of (a) sub- and standard-Chandrasekhar progenitor B-WDs, (b) standard- and
super-Chandrasekhar progenitor B-WDs and (c) sub- and super-Chandrasekhar progenitor B-WDs, via a single respective mass--radius relation
for these cases.

\section*{acknowledgments}
Research of DD is funded by the C.V. Raman Postdoctoral Fellowship (Reg. No. R(IA)CVR-PDF/2020/222) from the Department of Physics, Indian Institute of Science. BM acknowledges partial support by a project of the Department of Science and Technology (DST), India, with Grant No. DSTO/PPH/BMP/1946 (EMR/2017/001226). FW is supported through the U.S.\ National Science Foundation under Grant PHY-2012152. 

\software{Jupyter Notebook (\citealt{Kluyver2016}), Python 3 (\citealt{Dalcin2008,Van2009}) with the packages math, decimal and time, Numerical Python (numpy, \citealt{oliphant2006,van2011}), Matplotlib (\citealt{Hunter2007}), and Scientific Python (scipy, \citealt{Virtanen2020}).}

\end{document}